\documentclass[prb,twocolumn,superscriptaddress]{revtex4-2}
\usepackage{ulem}
\usepackage{graphicx,amsmath,amssymb,amsfonts,verbatim,dsfont,color}
\usepackage{bbm, bm, latexsym}
\usepackage{hyperref}
\usepackage[us]{datetime}
\usepackage{tikz,soul}
\usepackage{lmodern}

\graphicspath{ {pics/} }

\definecolor{lime}{HTML}{A6CE39}
\DeclareRobustCommand{\orcidicon}{%
	\begin{tikzpicture}
	\draw[lime, fill=lime] (0,0)
	circle [radius=0.16]
	node[white] {{\fontfamily{qag}\selectfont \tiny ID}};
	node[white] {{\fontfamily{cmr}\selectfont \tiny ID}}
	\draw[white, fill=white] (-0.0625,0.095)
	circle [radius=0.007];
	\end{tikzpicture}
	\hspace{-2mm}
}

\foreach \x in {A, ..., Z}{%
	\expandafter\xdef\csname orcid\x\endcsname{\noexpand\href{https://orcid.org/\csname orcidauthor\x\endcsname}{\noexpand\orcidicon}}
}

\newcommand{\kp}{$k\cdot p$ }

\newcommand{\thetitle}{Topological phase diagram and quantum magnetotransport effects in (Pb,Sn)Se quantum wells with magnetic barriers (Pb,Eu)Se}
\begin{document}

\title{\thetitle}

\author{Alexander~Kazakov\orcidA}
\email{kazakov@MagTop.ifpan.edu.pl}
\affiliation{International Research Centre MagTop, Institute of Physics, Polish Academy of Sciences, Aleja Lotnikow 32/46, PL-02668 Warsaw, Poland}
\author{Valentine~V.~Volobuev\orcidH}
\email{volobuiev@MagTop.ifpan.edu.pl}
\affiliation{International Research Centre MagTop, Institute of Physics, Polish Academy of Sciences, Aleja Lotnikow 32/46, PL-02668 Warsaw, Poland}
\affiliation{National Technical University "KhPI", Kyrpychova Str. 2, 61002 Kharkiv, Ukraine}
\author{Chang-Woo~Cho\orcidC}
\affiliation{Laboratoire National des Champs Magn{\'e}tiques Intenses, CNRS, LNCMI, Universit{\'e} Grenoble Alpes, Universit{\'e} Toulouse 3, INSA Toulouse, EMFL, F-38042 Grenoble, France}
\affiliation{Department of Physics, Chungnam National University, Daejeon, 34134, Republic of Korea}
\author{Benjamin~A.~Piot\orcidD}
\affiliation{Laboratoire National des Champs Magn{\'e}tiques Intenses, CNRS, LNCMI, Universit{\'e} Grenoble Alpes, Universit{\'e} Toulouse 3, INSA Toulouse, EMFL, F-38042 Grenoble, France}
\author{Zbigniew~Adamus\orcidB}
\affiliation{Institute of Physics, Polish Academy of Sciences, Aleja Lotnikow 32/46, PL-02668 Warsaw, Poland}
\author{Tomasz~Wojciechowski\orcidE}
\affiliation{International Research Centre MagTop, Institute of Physics, Polish Academy of Sciences, Aleja Lotnikow 32/46, PL-02668 Warsaw, Poland}
\author{Tomasz~Wojtowicz\orcidG}
\affiliation{International Research Centre MagTop, Institute of Physics, Polish Academy of Sciences, Aleja Lotnikow 32/46, PL-02668 Warsaw, Poland}
\author{Gunther~Springholz\orcidF}
\affiliation{Institut f{\"u}r Halbleiter- und Festk{\"o}rperphysik, Johannes Kepler University, Altenbergerstrasse 69, A-4040 Linz, Austria}
\author{Tomasz~Dietl\orcidI}
\email{dietl@MagTop.ifpan.edu.pl}
\affiliation{International Research Centre MagTop, Institute of Physics, Polish Academy of Sciences, Aleja Lotnikow 32/46, PL-02668 Warsaw, Poland}

\begin{abstract}
In this study, we report here on a successful growth by molecular beam epitaxy of high crystalline quality Pb$_{1-x}$Sn$_{x}$Se:Bi/Pb$_{1-y}$Eu$_{y}$Se QWs with $x = 0.25$ and $y = 0.1$, and on their magnetotransport characterization as a function of the QW thickness between 10 and 50\,nm, temperatures down to 300\,mK, perpendicular and tilted magnetic fields up to 36\,T. The character of weak antilocalization magnetoresistance and universal conductance fluctuations points to a notably long phase coherence length. It is argued that a relatively large magnitude of the dielectric constant of IV-VI compounds suppresses the decoherence by electron-electron scattering. The observation of Shubnikov-de-Haas oscillations and the quantum Hall effect, together with multiband \kp modelling, have enabled us to assess valley degeneracies, the magnitude of strain, subbands effective masses, and the topological phase diagram as a function of the QW thickness. Our results demonstrate that further progress in controlling Sn content, carrier densities, and magnetism in Pb$_{1-x}$Sn$_{x}$Se/Pb$_{1-y}$Eu$_{y}$Se QWs will allow for the exploration of the topologically protected quantized edge transport even in the absence of an external magnetic field.
\end{abstract}

\date{\today}
\maketitle

\section{Introduction}

In recent decades, the quantum Hall effects and other quantum transport phenomena have been observed in a wide range of topological materials, including topological insulators, Dirac semimetals, and Weyl semimetals \cite{Zhang2005,Buettner2011,Xu2014,Qiu2020}. However, one notable exception in this series is the topological crystalline insulators (TCIs), in which the presence of topological surface and interfacial 2D states was demonstrated by angle-resolved photoelectron spectroscopy (ARPES) \cite{Dziawa2012,Tanaka2012,Xu2012} and magnetooptical studies \cite{Krizman2022} but there is lack experimental evidence for quantized edge resistance in either weak or high magnetic fields. In TCIs, the topological protection relies on the crystal point group symmetries \cite{Fu2011}. The archetypical example of a TCI is SnTe \cite{Hsieh2012}, although this topological phase can also be achieved in other cubic IV-VI monochalcogenides with appropriately tuned compositions. The band structure of IV-VI semiconductors gives rise to distinct quantum Hall (QHE) features in these compounds \cite{Chitta2005,Pena2021,Krizman2024}, such as an unconventional sequence of filling factors $\nu$ \cite{Chitta2005} or the strain-controlled valley polarization in the 2D system \cite{Krizman2024}. The development of a robust TCI-based platform for quantum spin Hall (QSH) system would also allow experimentally access to the long-sought theoretical predictions, including the large-Chern-number quantum anomalous Hall effect (QAHE) \cite{Fang2014,Niu2015}, topological transistor \cite{Liu2013}, and other phenomena unique to the thin TCI layers \cite{Tang2014,Li2016a}. Also, recently IV-VI semiconductors have been considered an attractive platform for realizing non-Abelian excitations \cite{Kate2022,Gomanko2022,Schellingerhout2023,Song2023}. These advances could open new avenues for research and applications in the field of topological quantum materials. 

\begin{figure}
    \centering
    \includegraphics[width=\columnwidth]{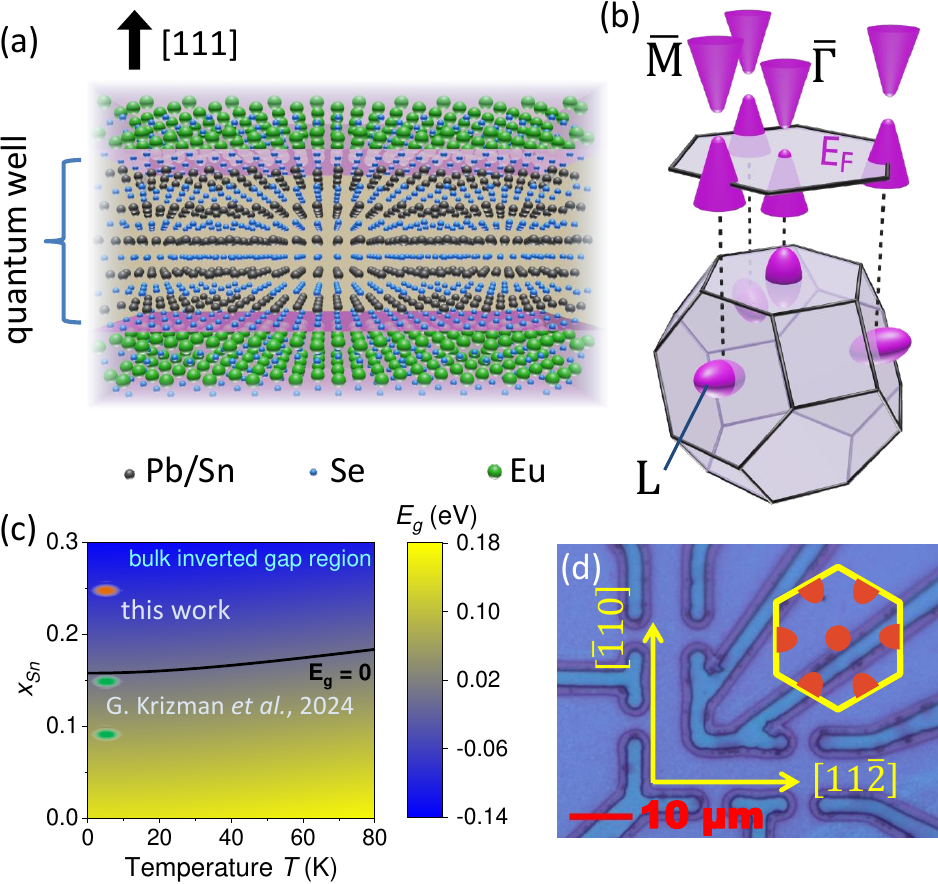}
    \caption{(a, b) A schematic representation of the grown QWs and their corresponding band structure showing four equivalent ellipsoidal valleys at the $L$ points. Those ellipsoids are projected onto the (111) surface at the $\overline{\Gamma}$ and $\overline{M}$ points. The Fermi level $E_{\mathrm{F}}$ crosses the $\overline{M}$ bands and $\overline{\Gamma}$ band at different levels. The energy offset between the $\overline{M}$ and $\overline{\Gamma}$ bands is due to strain induced by the lattice mismatch between the QW and barrier materials. (c) The bulk band gap of Pb$_{1-x}$Sn$_{x}$Se as a function of temperature $T$ and tin content $x_{\mathrm{Sn}}$. The samples studied in the current work fall within the region with an inverted gap (blue region above the black line indicating the zero gap), whereas previous QHE studies \cite{Krizman2024} focused on samples with lower tin content. (d) An optical image of two Hallbars oriented along the $[11\overline{2}]$ and $[\overline{1}10]$ directions, respectively. This design enables the observation of potential effects due to the anisotropic band structure, with its relative orientation schematically illustrated in the inset.}
    \label{fig:bands}
\end{figure}

The choice of crystal orientation plays a critical role in determining the nature of surface states in TCIs \cite{Liu2013a}. Specifically, the growth orientation of a TCI-based 2D system defines the character of zero-field quantized edge transport \cite{Safaei2015,Liu2015}. The most common growth orientations are $(100)$ and $(111)$, which can be used for the realization of the 2D TCI phase or QSH, respectively. A typical substrate for epitaxial growth of IV-VI semiconductors is $(111)$~BaF$_2$. As a result, in the 2D $(111)$ surface projection of band structure, the valence and conduction bands are close to each other at the $\overline{\Gamma}$ point (commonly referred to as the longitudinal valley) and at the three inequivalent $\overline{M}$ points (oblique valleys), as illustrated in Fig.~\ref{fig:bands}. Pb$_{1-x}$Eu$_{x}$Se(Te) is the preferred barrier material for Pb$_{1-x}$Sn$_{x}$Se(Te) quantum well (QW) heterostructures, due to its closely matched lattice constants and thermal expansion coefficients. However, a residual lattice mismatch between the QW and the barrier materials affects the band gap and generates an energy offset between $\overline{\Gamma}$ and $\overline{M}$ bands \cite{Simma2012,Simma2014}. This offset is responsible for the unconventional sequence of QHE filling factors \cite{Chitta2005} and plays a key role in controlling valley polarization in these systems \cite{Krizman2024}.

Previous magnetotransport studies of IV-VI QWs \cite{Springholz1993,Olver1994,Chitta2005,Pena2021,Krizman2024} have focused on samples with low Sn content, where the band inversion does not occur in the bulk crystal, and the TCI phase is absent. In this work, we report on a successful growth by molecular beam epitaxy of high crystalline quality Pb$_{0.75}$Sn$_{0.25}$Se:Bi/Pb$_{0.9}$Eu$_{0.1}$Se QWs with the thicknesses ranging from 10 to 50\,nm. The achieved Sn content $x = 0.25$ corresponds to the TCI phase in the bulk crystal. The Sn content achieved here is significantly higher than in the previously studied (Pb,Sn)Se QWs that exhibit the quantum Hall effect \cite{Krizman2024}. Typically, a higher level of Sn alloying results in a lowering of carrier mobility and increasing hole density \cite{Wang2023}, which is challenging to control even with Bi doping \cite{Volobuev2017}. Our results demonstrate that for the developed growth protocols, the larger Sn content is not accompanied by a significant increase in the carrier concentration and lowering crystal quality, so that the quantum Hall effect is observed along with pronounced weak antilocalization magnetoresistance and universal conductance fluctuations owing to a long phase coherence length at low temperatures. Finally, our findings show to what extent persistent photoconductivity can serve to control the carrier density in (Pb,Sn)Se QWs.

Longitudinal and Hall resistances of the patterned QWs have been studied at temperatures down to 300\,mK and in perpendicular and tilted magnetic fields up to 36\,T. In weak magnetic fields, we observe and interpret theoretically weak antilocalization (WAL) magnetoresistance and universal conductance fluctuations (UCF). We find that a relatively large magnitude of the dielectric constant results in weak decoherence by carrier-carrier scattering in IV-VI compounds, which enlarges WAL magnetoresistance and makes it possible to detect UCF in our 5 $\mu$m-size Hall bars. The observation of Shubnikov-de-Haas (SdH) oscillations and the quantum Hall effect (QHE), together with multiband \kp modelling, has allowed us to assess valley degeneracies, the magnitude of strain, subbands effective masses, and the topological phase diagram as a function of the QW thickness. We argue that a large magnitude of the dielectric constant, resulting in a double occupation of localized states, may reduce scattering between helical edge states compared to the case of HgTe-type topological insulators \cite{Dietl2023,Dietl2023a}. In general terms, our results demonstrate that further progress in controlling Sn content, carrier densities, strain, and magnetism in Pb$_{1-x}$Sn$_{x}$Se QWs will allow for the exploration of topologically protected quantized edge transport even in the absence of an external magnetic field with a prospect for quantum metrology applications.

\section{Samples and experimental details}

Following previous developments for IV-VI systems \cite{Krizman2024,Kazakov2021,Turowski2023}, molecular beam epitaxy (MBE) of Pb$_{0.75}$Sn$_{0.25}$Se/Pb$_{0.9}$Eu$_{0.1}$Se QWs studied here has been carried out on (111) BaF$_{2}$ substrates using a Riber 1000 MBE system. The growth process was conducted in the ultra-high vacuum environment with a base pressure of $<5\times10^{-10}$~mbar, utilizing compound PbSe and SnSe and elemental Eu and Se as the source materials. The composition of the QWs and barrier layers is determined by the beam flux ratio (SnSe/PbSe and PbSe/Eu), which is controlled by a quartz microbalance positioned at the substrate location. The substrate temperature was maintained at 350~$^\circ$C, and the growth rate was approximately 1~$\mu$m/hour. The Pb$_{0.9}$Eu$_{0.1}$Se buffer and cap layers were grown to thicknesses of 2--3~$\mu$m and 200~nm, respectively. To achieve low carrier densities and counterbalance the natural hole doping caused by Sn vacancies, extrinsic n-type Bi-doping (nominally 0.004\%) was introduced using a Bi$_{2}$Se$_{3}$ doping cell. The sample surface quality is monitored {\it in-situ} using reflection high-energy electron diffraction (RHEED). Post-growth analyses, including atomic force microscopy (AFM) and X-ray diffraction (XRD), confirmed the high crystalline quality of the grown structures. 

\begin{figure}
    \centering
    \includegraphics[width=\columnwidth]{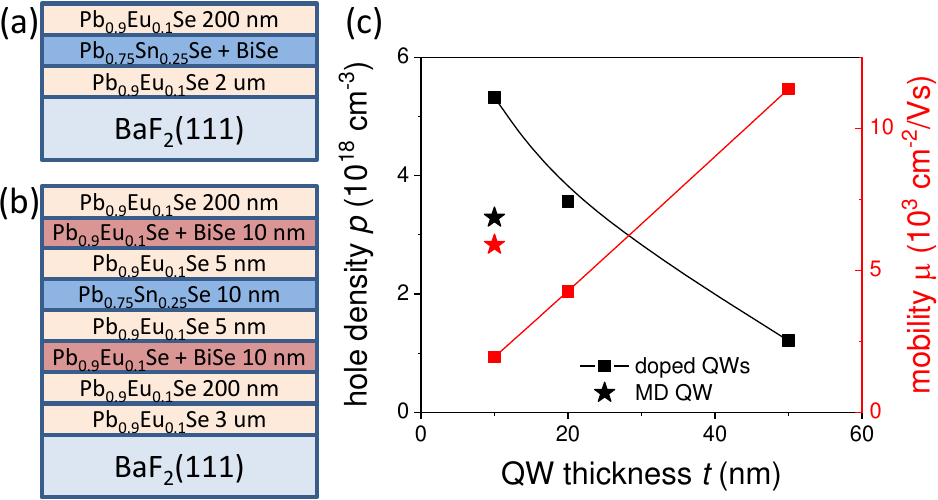}
    \caption{Layout of structures with Bi doping in the wells (a) and in the barriers (b). Carrier density and mobility plotted as a function of QW thickness (c). The square and star mark two different doping designs depicted in (a) and (b), respectively.}
    \label{fig:layout}
\end{figure}

The current study focused on three Bi-doped QWs with thicknesses of 50, 20 and 10~nm, referred to as samples A, B and C, respectively, whose layout is shown in Fig.~\ref{fig:layout}(a). A fourth sample (labelled as D), a 10\,nm QW is modulation-doped, i.e., contains Bi-doped layers in the barriers, 5~nm away from the QW, as depicted in Fig.~\ref{fig:layout}(b). Preliminary magnetoresistance and Hall effect characterization of the unprocessed samples, obtained in a home-built 1.5~K/9~T setup, can be found in \cite{sup}. The low-field Hall slope and zero field resistivity indicate relatively low hole density, in the $10^{18}$~cm$^{-3}$ range and the mobility values between 2$\times10^{3}$ and 11$\times10^{3}$~cm$^{2}$/Vs, as summarized in Table~\ref{tab:info} and in Fig.~\ref{fig:layout}. In general, carrier mobility decreases with QW thickness, while the 2D carrier density remains nearly constant. Notably, a decrease in carrier density is also accompanied by increased mobility, as can be noted by comparing samples~C and D.

\begin{table*}
\caption{Nominal Sn content $x_{\text{Sn}}$, QW thickness $d_{\text{QW}}$, Bi doping $x_{\text{Bi}}$, bulk carrier density $p_{3D}$, and corresponding sheet carrier density $p_{2D}$, and carrier mobility $\mu$ for unprocessed samples; concentrations $p^{\text{Hall}}$ and $p^{\text{SdH}}$, and valley degeneracy $n_v$ are for Hall bars. Here we also specified values of the thermal ($l_{T}$) and phase coherence ($l_{\phi}$) lengths at the lowest temperature, which is relevant for the WAL and UCF analysis.}
\begin{tabular}{c|c c c c c c c c c c c c}
\hline
QW & $x_{\text{Sn}}$ & $d_{\text{QW}}$ & $x_{\text{Bi}}$& $p_{3D}$ & $p_{2D}$, & $\mu$& $p^{\text{Hall}}$& $p^{\text{SdH}}$& $n_v$ & $l_{T}$(300~mK) & $l_{\phi}$(300~mK)\\
 &           \%      & nm   &\% &$10^{18}$~cm$^{-3}$ &$10^{12}$~cm$^{-2}$ &cm$^2$/Vs& $10^{12}$~cm$^{-2}$ & $10^{12}$~cm$^{-2}$ & & $\mu$m & $\mu$m\\
\hline
A & 24 & 50 & 0.004 & 1.21 & 6.03 & 11400& 4.50 & 4.08 & 2 $\cdot$ 3 & 1 & 6.0\\
B & 25 & 20 & 0.004 & 3.56 & 7.13 & 4250 & 7.05 & 4.35 & 3 & 0.9 & 7.9\\
C & 25 & 10 & 0.0042 & 5.32 & 5.32 & 1950 & 5.52 & 5.16 & 3 & 0.5 & 2.0\\
D (MD) & 25 & 10 & 0.004 & 3.29 & 3.29 & 5900 & 3.28 & 2.82 & 3 & 0.7 & 3.0\\
\hline
\end{tabular}
\label{tab:info}
\end{table*}

The chosen heterostructures have been processed into L-shaped Hall bar structures (see Fig.~\ref{fig:bands}(d)), using optical lithography and wet Br etching techniques. One arm of the Hall bar is aligned along the $[11\overline{2}]$ crystallographic direction, while the other --- along the $[\overline{1}10]$ direction. This configuration allows for a comparison of transport properties when the current is applied either perpendicular or parallel to one of the oblique valleys. These structures were initially studied in moderate fields (up to 7~T) in the He$^3$ cryostat, equipped with a piezoelectric rotator. Later, the same samples were investigated in high magnetic fields (up to 36~T) in the Grenoble High Magnetic Field Laboratory in a pumped helium cryostat, equipped with a manual rotator. In both setups, the rotators allow for the magnetic field tilt angle between 0$^{\circ}$ and 90$^{\circ}$, the perpendicular and in-plane configuration, respectively.

Table~\ref{tab:info} shows the hole density $p^{\text{Hall}}$ determined for Hall bars from the low-field Hall resistance slope, $dR_{xy}/dB = 1/ep^{\text{Hall}}$ compared to the value obtained from SdH oscillations, $p^{\text{SdH}}= n_v(e/h)(F_1+F_2)$, where $F_{1,2}$ are the positions of the FFT peaks and the valley degeneracy $n_v= 3$ for thinner QWs, as there are three oblique valleys occupied. For sample~A, to account for spin degeneracy, we use $n_v=6$, as the spin splitting is not resolved in the FFT spectra. Overall, we can see that the carrier densities determined by both methods are consistent within approximately $\approx10$~\%, except for sample~B. This discrepancy may be due to the presence of either a second subband or longitudinal valley at $E_{F}$. Such factors could also explain the pronounced magnetoresistance and the low amplitude of SdH oscillations observed at low magnetic fields for this sample \cite{sup}.

\section{Results and Discussion}
\subsection{\kp modelling and topological phase diagram}
\label{sec:kp1}

We have carried out computations of electric subbands and Landau levels in $\langle 111\rangle$ Pb$_{0.75}$Sn$_{0.25}$Se/Pb$_{0.9}$Eu$_{0.1}$Se QWs employing a four band \kp model \cite{Mitchell1966,Nimtz1983,Bauer1992} taking into account the presence of confinement, strain, exchange interactions with Eu spins, and external magnetic field. We assume that, except for the gap, the band structure parameters are independent of Sn and Eu content $x$ and $y$. Further details about the \kp Hamiltonian and computational method are provided in \hyperref[app:kp]{Appendix}.

\begin{figure*}
    \centering
    \includegraphics[width=2.00\columnwidth]{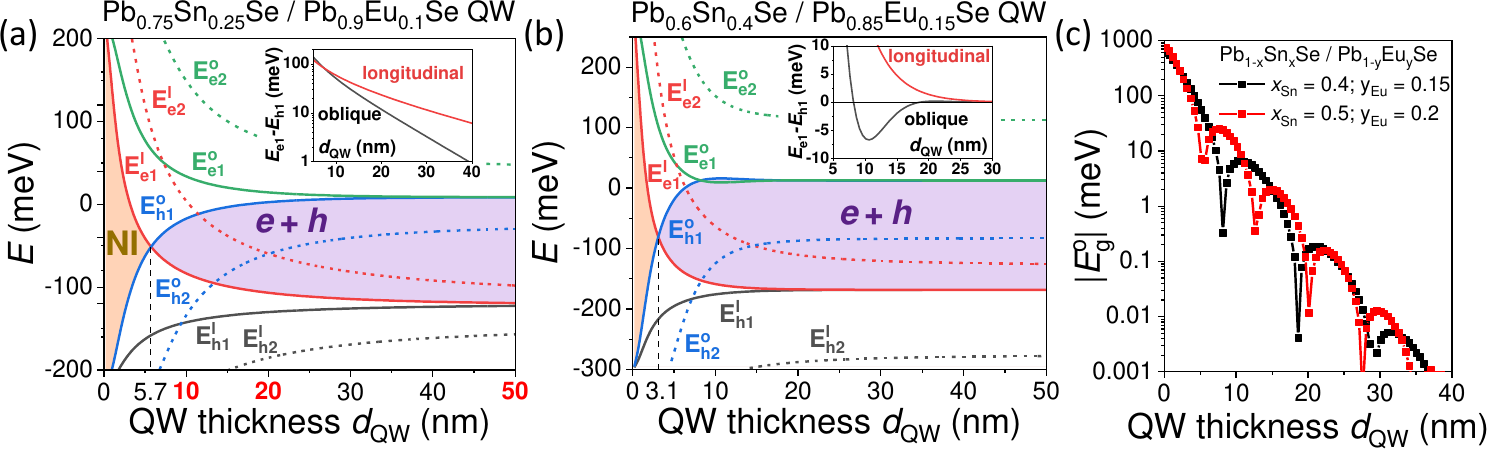}
    \caption{(a) First two subbands for electrons and holes in oblique and longitudinal valleys (solid and dashed lines stand for the first and second subbands respectively) are plotted as a function of thickness for the case of Pb$_{0.75}$Sn$_{0.25}$Se/Pb$_{0.9}$Eu$_{0.1}$Se QW. Thicknesses $d_{QW}$ of 50, 20 and 10~nm on this diagram correspond to the samples A, B, C and D. (b) The same for a higher tin content QW -- Pb$_{0.6}$Sn$_{0.4}$Se/Pb$_{0.85}$Eu$_{0.15}$Se QW. The insets in (a,b) show the corresponding distance between the first electron and hole subbands, $E_{e1}-E_{h1}$ that is always positive for both types of valleys in (a), but changes sign for oblique valleys in (b). (c) Oscillatory behaviour for the oblique valley gap for two different QW compositions.}
    \label{fig:kp_subbands}
\end{figure*}

By plotting energy subbands as a function of QW thickness $d_{\text{QW}}$ for various compositions, we can infer the transport properties for different compositions, thicknesses, and Fermi levels. For the composition used in the current work -- Pb$_{0.75}$Sn$_{0.25}$Se/Pb$_{0.9}$Eu$_{0.1}$Se, shown in Fig.~\ref{fig:kp_subbands}(a) -- we see that below $\approx$~5.7~nm, the system behaves a true insulator (shaded with orange). However, above this critical thickness, at the appropriate Fermi level, the system enters a mixed conduction state, with the contribution of both electrons and holes that originate from different valleys, as described in \cite{Krizman2024}. Although the bulk gap is inverted, in the heterostructure, the electron subbands $E^{l,o}_{e1}$ are always above hole subbands $E^{l,o}_{h1}$ for the corresponding valleys for reasonable $d_{\text{QW}}$ values. The second subbands $E^{l,o}_{e2,h2}$ asymptotically tend to the values close to the bulk band edges at higher thicknesses.

The situation changes when we consider QWs with higher Sn content. For a QW with the composition Pb$_{0.6}$Sn$_{0.4}$Se/Pb$_{0.85}$Eu$_{0.15}$Se, Fig.~\ref{fig:kp_subbands}(b), we clearly see the region where $E^{o}_{h1}$ goes above $E^{o}_{e1}$. This indicates the gap inversion that occurs in the heterostructure and was previously described for QWs with higher Sn compositions \cite{Safaei2015,Liu2015,Rechcinski2021}. However, in the current work, we consider this problem with a model based on experimental results and in a realistic environment. The current result agrees with previous works, as it also demonstrates: (i) the damped oscillatory behaviour of the gap in oblique $\overline{M}$ valleys \cite{Safaei2015,Liu2015}, (ii) the absence of the gap inversion in the longitudinal valley \cite{Safaei2015,Liu2015}, and (iii) changes in the oscillation period with varying Sn content (Fig.~\ref{fig:kp_subbands}(c)) \cite{Rechcinski2021}. Thus, the employed \kp model can quantitatively reproduce previous experimental results and also agree with previously published theoretical predictions obtained within the tight-binding approximation.

\begin{figure}[htb]
    \centering
    \includegraphics[width=0.80\columnwidth]{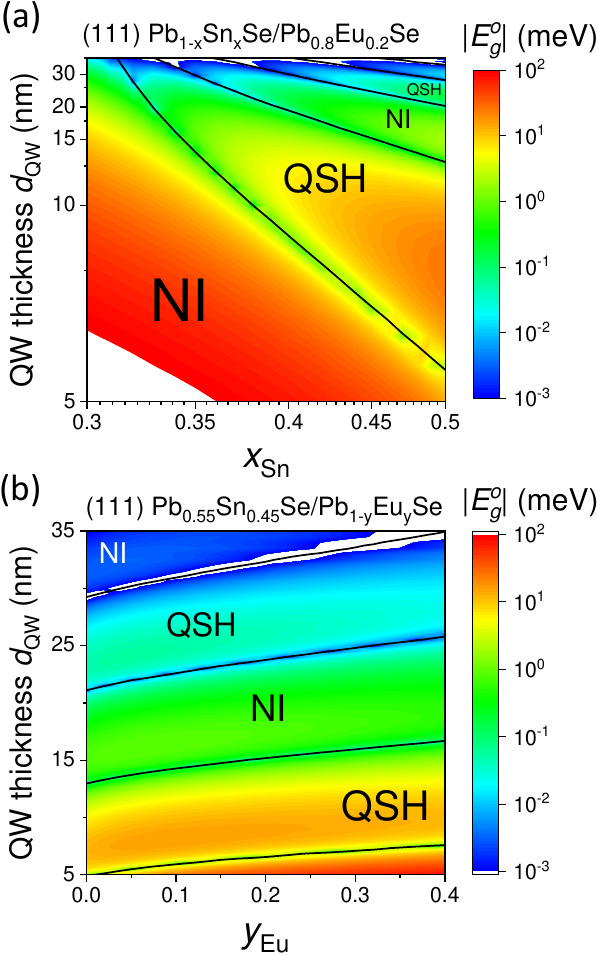}
    \caption{Alteration of normal insulator (NI) and quantum spin Hall (QSH) phases with QW thickness and composition (as a function of Sn content in the QW (a) and Eu content in the barriers (b)).}
    \label{fig:kp_th}
\end{figure}

Inset in figure \ref{fig:kp_subbands}(a) show the band gaps at the $\overline{\Gamma}$ and $\overline{M}$ points without an external magnetic field as a function of the QW thickness $d_{\text{QW}}$ for the Sn and Eu content used in the QWs studied in the current work. As expected, the confinement increases the band gap. Accordingly, the inverted order of bands, usually denoting the topological phase, is shifted to higher $x$ in QWs compared to the bulk case, from $x_{c}^{bulk}\approx0.16$ to $x_{c}^{QW}=0.31-0.50$ depending on the value of $d_{QW}$ (see Fig.~\ref{fig:kp_th}a). These results imply that our QWs are in the normal phase and should not show edge transport by helical states.

The described \kp model can also be utilized to design TCI-based heterostructures to access topological edge transport. For instance, we can calculate the variation of the energy gap between $E_{h1}^{o}$ and $E_{e1}^{o}$ as a function of the Sn content in the QW (Fig.~\ref{fig:kp_th}(a)) or as a function of Eu content in the barriers (Fig.~\ref{fig:kp_th}(b)). Please note that in Fig.~\ref{fig:kp_th}(a) QW thickness is presented on a reciprocal scale, while the Sn content is on a logarithmic scale. Over a wide range of parameters, the lines delimiting the gap inversion are very close to linear in these scales, providing a practical tool for evaluating the likelihood of observing gap inversion effects. However, accumulating more experimental data would help refine the values of the band parameters (Table~\ref{tab:bandparam}) used in the model and better understand their dependence on composition. Additionally, a self-consistent expansion of the current model would enhance the precision of the results \cite{Beugeling2025}.

\subsection{Weak antilocalization effect and universal conductance fluctuations }
\label{sec:WAL}

\begin{figure*}
    \centering
    \includegraphics[width=2\columnwidth]{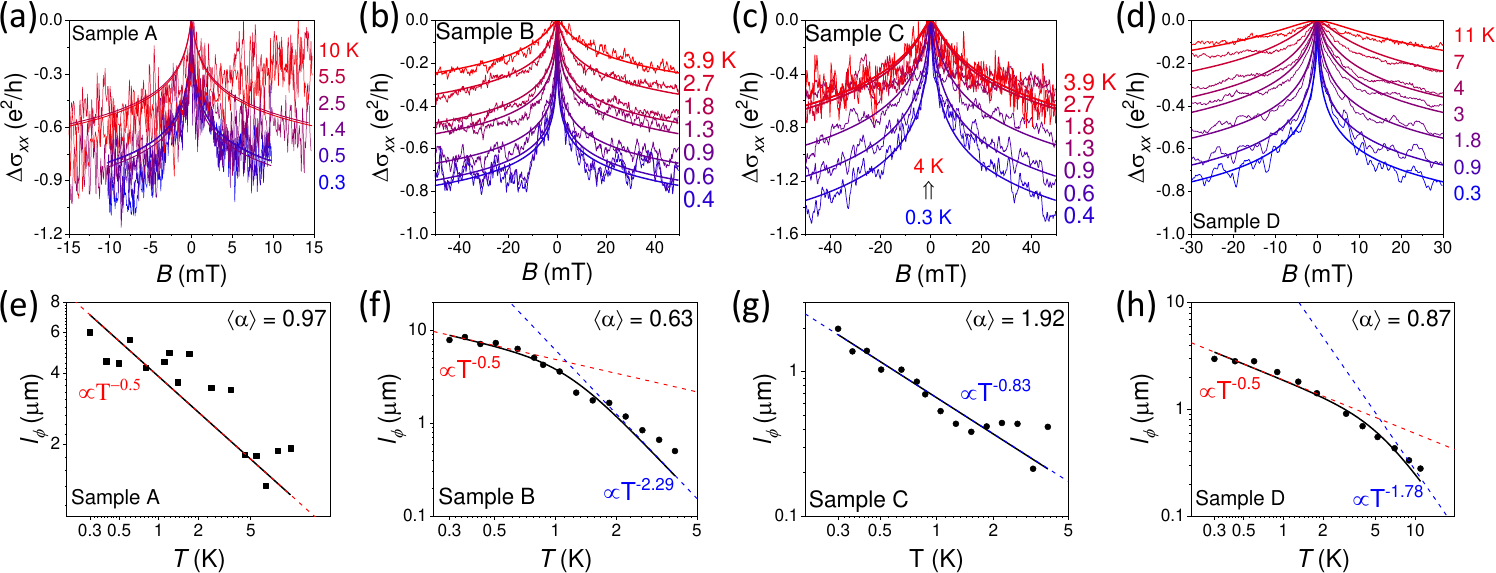}
    \caption{(a-d) Magnetoconductance at various temperatures (thin lines) showing the presence of WAL and UCF effects, and the corresponding fit to the HLN expression \ref{eq:HLN} (thick lines). (e-h) Values of the fitted prefactor $\langle\alpha\rangle$ and phase coherence lengths $l_\phi(T)$ pointing to dephasing by carrier-carrier and phonon scattering at low and high temperatures, respectively.}
    \label{fig:WAL}
\end{figure*}

Figure~\ref{fig:WAL} shows a pronounced, temperature dependent cusp in the negative weak-field magnetoconductance $\Delta \sigma_{xx}(B)$ indicating the presence of the WAL effect, as found previously in thin epilayers \cite{Kazakov2021} and QWs \cite{Wang2020} of Pb$_{1-x}$Sn$_{x}$Se and other narrow-gap semiconductors \cite{Dietl1993}. Additionally, aperiodic and reproducible fluctuations of $\Delta \sigma_{xx}(B)$, known as UCF, are also clearly visible in our Hall bridges with about 5 to 10\,$\mu$m distance between the contact probes. 

We begin by describing the WAL effect employing the Hikami-Larkin-Nagaoka (HLN) expression for the case of strong SOC \cite{Hikami1980}, 
\begin{equation}
    \Delta \sigma_{xx}(B) = -\alpha \frac{e^2}{2\pi^{2}\hbar} \left[ \psi \left( \frac{1}{2} + \frac{B_{\phi}}{B} \right) - \ln \left( \frac{B_{\phi}}{B} \right) \right],
    \label{eq:HLN}
\end{equation}
where the prefactor $\alpha = n_v/2$, where $n_v$ is a number of valleys uncoupled by inter-valley scattering; $B$ is the magnetic field perpendicular to the sample plane; $\psi(x)$ is the digamma function; and $B_\phi=\hbar/4el_\phi^2$ is the dephasing field determined by the phase coherence length $l_\phi(T)$. We start by fitting $\Delta \sigma_{xx}(B)$ at the lowest temperature $0.5 \ge T \ge 0.3$\,K treating $\alpha(T)$ and $l_\phi(T)$ as two adjustable parameters, though the presence of large fluctuations somewhat deteriorates the fit accuracy. With $\langle\alpha\rangle$ values obtained as an average of $\alpha(T)$ over that narrow temperature range, we determine $l_\phi(T)$ in the whole studied temperature range up to 13\,K. The observed values of $l_{\phi}$ in IV-VI materials systems at low temperatures are among the longest reported in the literature \cite{Kazakov2021,Wang2023}. However, the $l_{\phi}$ magnitudes decrease with temperature, so that spin-orbit length $l_{SO}$ ceases to be the shortest lengthscale at high temperatures, resulting in a discrepancy between the experimental data and the HLN fitting in the strong-SO limit noticeable in Fig.~\ref{fig:WAL}. As shown in Figs.~\ref{fig:WAL}(e-h), $\langle\alpha\rangle$ magnitudes varies between 0.4 to almost 2 for $l_\phi(0.3\,\mbox{K})$ between 8 and 2\,$\mu$m, respectively. Such correlation between $\langle\alpha\rangle$ and $l_\phi$ may indicate that intervalley scattering reduces $n_v$ to 1 in QWs with sufficiently large magnitudes of $l_\phi$. Furthermore, temperature dependencies of $l_\phi(T)$ summarized in Figs.~\ref{fig:WAL}(e-h), suggest the presence of two phase breaking mechanisms. Accordingly, we describe $l_\phi(T)$ by the formula,
\begin{equation}
    l_{\phi}=\left(\frac{T}{l_0^2}+\frac{T^{p}}{l^2}\right)^{-1/2},
\end{equation}
which takes into account dephasing by carrier-carrier scattering in the form expected for the 2D case and the electron-phonon dephasing process dominating at higher temperatures, $p>1$. A relatively minor importance of dephasing by carrier-carrier scattering in IV-VI compounds results, as already noted \cite{Prinz1999,Kazakov2021,Wang2020,Geng2022}, from a sizable magnitude of the dielectric constant in systems close to a ferroelectric instability \cite{Grabecki2007,Kolwas2013}. Long dephasing lengths provide evidence for the suppressed decoherence processes, which is beneficial for the achievement of longer QSH lengthscales \cite{Lunczer2019} and possible applications in quantum technologies \cite{Islam2022}. Though a particular mechanism affecting the violation of the topologically quantized edge transport can be material dependent, any possibility to suppress one of the probable scattering channels and achieve longer helical edge channels is valuable. Conclusions about decoherence length can hardly be drawn from the magnetooptical spectra or the quantum Hall effect in samples with low Sn content.

Our previous studies of 50-nm-thick $p$-type Pb$_{1-x}$Sn$_{x}$Se epilayers across the topological phase transition \cite{Kazakov2021} indicated that rather than the Rashba effect \cite{Peres2014}, spin-orbit locking (or the Elliott-Yafet mechanism in the time honored nomenclature) is a dominant mechanism accounting for WAL magnetoresistance, as no Rashba splitting of bands has been detected in the ARPES spectra \cite{Kazakov2021,Krizman2024}. We will return to this issue discussing beating of SdH oscillations.

\begin{figure}
    \centering
    \includegraphics[width=0.65\columnwidth]{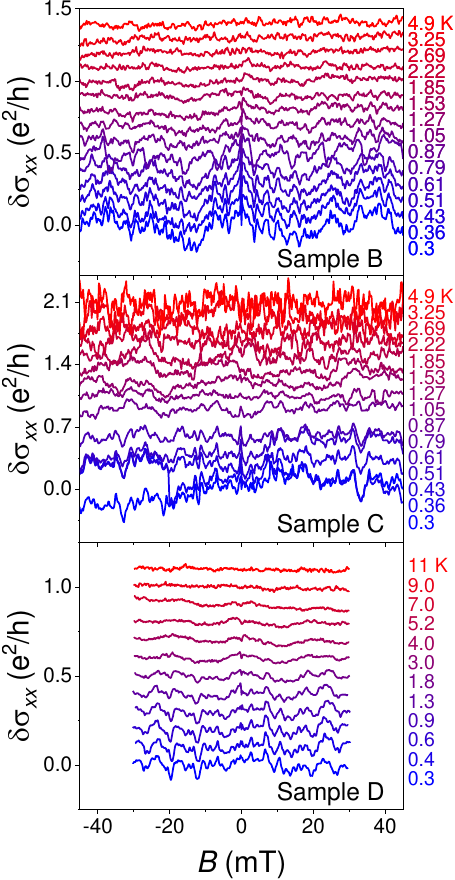}
    \caption{UCF's in the studied samples after subtraction HLN fitting from the magnetoconductance data at different temperatures.}
    \label{fig:UCF}
\end{figure}

Turning to UCF, we show in Fig.~\ref{fig:UCF} magnetoconductivity data after subtracting the WAL background, $\delta\sigma_{xx}= \Delta\sigma_{xx}-\sigma_{\text{HLN}}^{\text{fit}}$. Relatively large magnitudes of $l_\phi(T)$ mean that the UCF magnitude is controlled by the thermal diffusion length \cite{Altshuler1986,Lee1987}, $l_T = (\hbar D/k_BT)^{1/2}$, where in the 2D case $D = 2\pi\hbar^2 p\mu/en_vm^*$ is the diffusion coefficient, the values of $p$ and $\mu$ are given Table~\ref{tab:info}, $n_v = 3$, and $m^*\approx0.15m_0$ according to SdH data discussed in the next section. For the determined parameters, we see that for all studied samples at the lowest temperature the $l_{T}<l_{\phi}$ and varies in the 0.5--1~$\mu$m range (Table~\ref{tab:info}).

From the low-temperature amplitude of the UCF shown in Fig.~\ref{fig:UCF}, which is $\delta\sigma^{\text{RMS}}\approx$ 0.07, 0.14 and 0.04~$e^2/h$ for samples~B, C and D respectively, we can estimate \cite{Altshuler1986,Lee1987} the $l_T$ value from $\delta\sigma_{xx}=0.862(e^2/h)\sqrt{3n_vs^2l_{T}^2/\beta wl}$, where $w$ and $l$ are the linear dimensions of the Hall bar, typically 5 and 10~$\mu$m's, respectively (Fig.~\ref{fig:bands}(d)). Assuming $n_v = \langle\alpha\rangle/0.5$, spin degeneracy $s = 2$ and the symplectic ensemble $\beta = 4$ in weak magnetic fields or $n_v = 2\langle\alpha\rangle/0.5$, $s = 1$ and $\beta = 2$, i.e., the unitary ensemble in stronger fields we find $l_T\approx$ 0.27, 0.35 and 0.13~$\mu$m at 0.3\,K for samples~B, C and D respectively, which is by a factor of two smaller than expected from the value of $D$. The value of $l_T$ can also be estimated from the half-width $B_{1/2}$ at the half-height of the autocorrelation function $C(\Delta B)=\left< \delta\sigma(B)\delta\sigma(B+\Delta B)\right>$, which is expected to be \cite{Lee1987} $B_{1/2}\approx h/2\pi el_{T}^2$. Using this method, we obtain low-temperature $l_T$ of 0.5, 0.8 and 0.9 for samples~B, C and D, respectively, which is close to those estimated from the value of $D$. 

\subsection{Quantizing Magnetic Fields}

\begin{figure}
    \centering
    \includegraphics[width=\columnwidth]{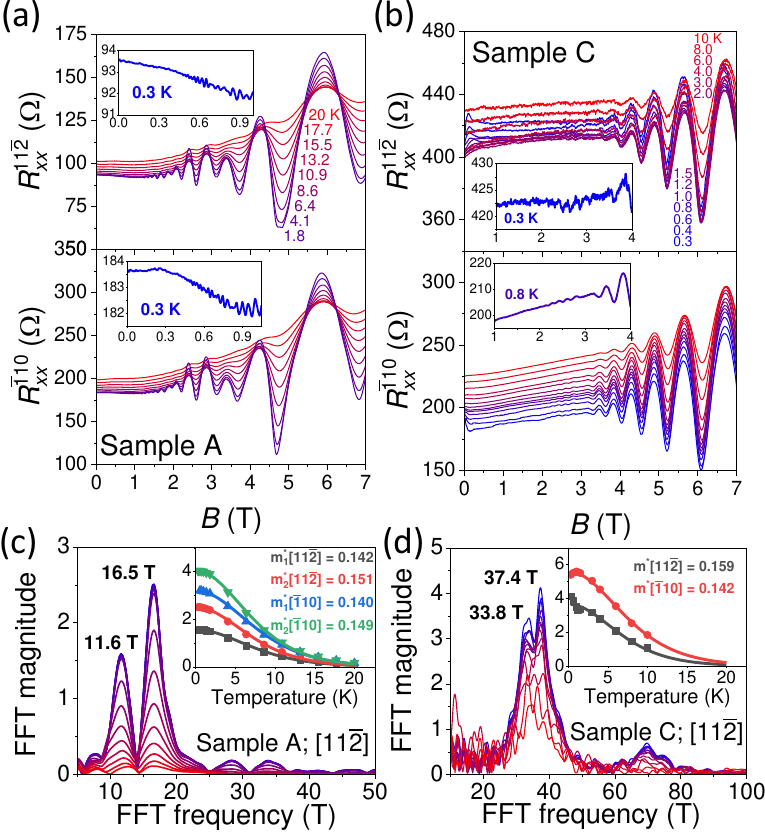}
    \caption{(a, b) SdH oscillations in samples~A and C, respectively in the temperature range from 13 to 0.6~K. Note that SdH oscillations measured along two perpendicular crystallographic axes display nearly identical patterns, suggesting that band anisotropy plays a minor role at relevant Fermi energies. (c, d) Temperature evolution of the FFT spectra for the SdH oscillations is shown in (a, b), respectively. Fitting of the temperature damping of the oscillations (eq.~\ref{eq:LK_therm}) in the insets provides values of the cyclotron masses. For the sample~A, values of the $m^*_{1}$ corresponds to the left peak, while $m^*_{2}$ -- to the right peak in the FFT spectra. For other samples, the values for the temperature damping fit were taken from the highest peak.}
    \label{fig:sdh}
\end{figure}

In the moderate field range, as shown in Figs.~\ref{fig:sdh}(a,b), we observe pronounced Shubnikov-de Haas (SdH) oscillations. They begin at approximately 0.5~T in the 50~nm-thick QW, indicating a large magnitude of the quantum lifetime $\tau_{\mathrm{q}}$ and, thus, pointing to a high quality of the heterostructures. As in other systems \cite{Dietl1978}, the values of quantum mobilities are lower than the weak-field Hall mobilities determined for the unprocessed QWs, shown in Table~\ref{tab:info}. As the QW thickness is reduced, both the quantum and Hall mobilities decrease significantly, underlying a crucial role of scattering at (Pb,Sn)Se/(Pb,Eu)Se interfaces. More specifically, the penetration of carriers' wavefunction into (Pb,Eu)Se barriers allows for sizable alloy scattering by Eu cations, which significantly reduces the mobility \cite{Prinz1999}. The Coulomb interaction with background and remote impurities is likely less effective due to a high magnitude of the static dielectric constant in IV-VI semiconductors \cite{Nimtz1983}. The oscillations' pattern is virtually identical for the current along the [11$\bar{2}$] and [$\bar{1}$10] directions.

A notable feature is the oscillations' beating pattern, suggesting the presence of two frequencies. Indeed, after subtracting a smooth background, the fast Fourier transformation (FFT) revealed two main frequencies for sample~A (Fig.~\ref{fig:sdh}(c,d)). However, for the sample~C, a splitting is found only below 1\,K. Similar splittings appear for two other samples \cite{sup}, which also correlate with the observation of the beating pattern. Spin-splitting persisting to low fields usually causes the beating pattern of the SdH oscillations. In our case, exchange interactions between carriers and barrier's Eu dopants can result in additional low-field spin splitting \cite{Teran2002} due to the penetration of the carrier's wavefunction into the barriers. However, such splitting should depend strongly on temperature, which is not confirmed by the data. Another source of spin splitting is spin-orbit coupling (SOC) and inversion asymmetry \cite{Hasegawa2003,Sandoval2022}, as detected in IV-VI QWs by photoelectron spectroscopy \cite{Rechcinski2021}. The observation of weak antilocalization (WAL) in the studied QWs, as discussed in Sec.~\ref{sec:WAL}, supports this interpretation.

We also note that the character of the beating patterns in thicker QW (A) and thinner QWs (B, C, D) is quite different. In thinner QWs we only see one node at around 2-3 T, while in sample A, we see several nodes. This is reflected in the corresponding FFT spectra, as in sample A, we see two well-separated peaks of different magnitudes, while in the thinner QW we observe only a slight doubling of the main peak. We interpret SdH beating in sample A by the presence of the second subband, as according to the \kp calculations the distance between the tops of the first and second oblique hole subbands decrease significantly with enlarging QW width, $E^{o}_{h1}-E^{o}_{h2}$ = 115, 62, and 38~meV for the 10 (samples C, D), 20 (B) and 50 nm (A) QWs, respectively (Fig.~\ref{fig:kp_subbands}a, blue curves).

The presence of SdH oscillations over a wide field range has made us possible to detect weak subband splitting in thinner samples. Two mechanisms can account for this observation: (i) residual in-plane strain that could partially lift threefold valley degeneracy or (ii) the Rashba SO effect, so far predicted in \cite{Hasegawa2003} and observed only in asymmetric IV-VI QWs \cite{Rechcinski2021}. In our case, a residual asymmetry can result from asymmetric terminations at the (Pb,Sn)Se/(Pb,Eu)Se interfaces or from interfacial electric fields, even though a large value of the dielectric constant should reduce their magnitude.

From temperature damping of the FFT peaks (Fig.~\ref{fig:sdh}(c,d)), we deduce the cyclotron mass of the carriers, using the temperature-dependent part of the Lifshitz-Kosevich equation \cite{Okazaki2018}:
\begin{equation}
    A_{\mathrm{FFT}}(T) = A_{0} \left( \frac{2\pi^2k_{\mathrm{B}}}{e\hbar} \frac{m^{*}T}{\bar{B}} \right) / \sinh\left( \frac{2\pi^2k_{\mathrm{B}}}{e\hbar} \frac{m^{*}T}{\bar{B}} \right),
    \label{eq:LK_therm}
\end{equation}
where $A_{0}$ is a fitting constant, $T$ is  temperature, $\bar{B}$ is the inverse of the mean value of the $1/B$ interval showing up in the FFT analysis, and $m^{*}$ is the cyclotron effective mass. Fitting the temperature dependence of the FFT peaks (insets in Figs.~\ref{fig:sdh}(c,d), values are given in units of electron mass $m_{0}$) yields $m^{*}\approx0.15m_{0}$ for the studied QWs, where $m_{0}$ is the free-electron mass. These values are close to the previously reported effective mass (0.13$m_{0}$) for $p$-type Pb$_{0.7}$Sn$_{0.3}$Se QW \cite{Wang2020}.

\begin{figure}
    \centering
    \includegraphics[width=\columnwidth]{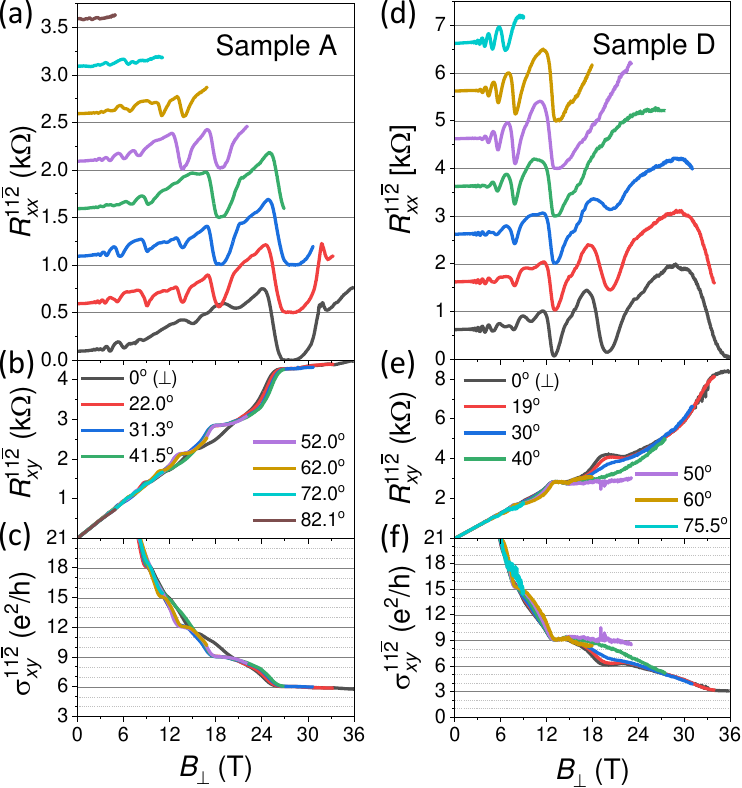}
    \caption{(a, d) Evolution of the SdH oscillations at 1.5~K with a tilt angle in high magnetic fields for the samples~A and D, respectively. For clarity, the curves are offset. In high fields $R_{xx}$ reaches almost zero value, which corresponds to plateaus of the R$_{xy}$ in (b, e). As the tilt angle increases, QH states at certain filling factors $\nu$ appear or disappear for both QWs. The $\sigma_{xy}$ plateaus are shown in (c, f), plotted in units of $e^{2}/h$. All the data is for the current along the $[11\overline{2}]$ direction.}
    \label{fig:qhe}
\end{figure}

High-field experiments also revealed the presence of the quantum Hall effect (QHE). An example of QHE plateaus is shown in Fig.~\ref{fig:qhe}. The observed QHE filling factors, $\nu$, are multiples of three, indicating that only oblique $\overline{M}$ valleys contribute to the QHE and the spin degeneracy is removed. It is interesting to note that according to results presented in Fig. 8, some QH plateaus only emerge at certain tilting angles. Within the simplest interpretation, such angles correspond to the situation in which the cyclotron and Zeeman splittings become equal. However, additional effects may come into play, such as lifting of valley degeneracy by the in-plane magnetic field \cite{Li2016a,Sodemann2017}.

\begin{figure}
    \centering
    \includegraphics[width=\columnwidth]{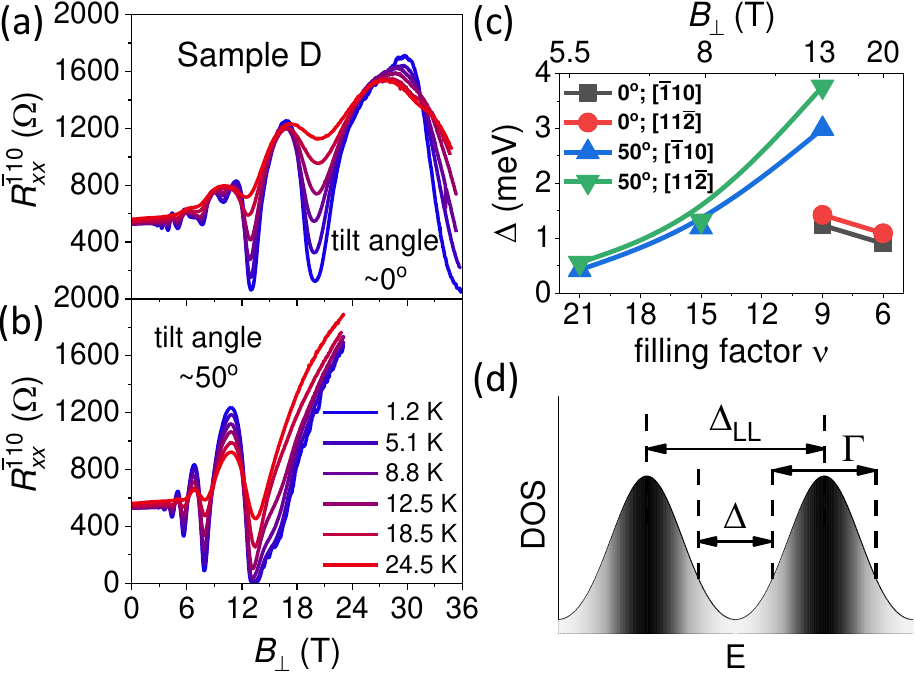}
    \caption{(a,b). Temperature evolution of the high field SdH oscillations and QHE states in the 50~nm QW for the two selected tilt angles. For the selected filling factors $\nu$, thermal activation energy gaps $\Delta$ (c) are determined. Comparison of those numbers with the distance between LL ($\Delta_{LL}$, Fig.~\ref{fig:kp}) points to the significant broadening $\Gamma$ of the LLs ({\bf e}) at high magnetic fields. (d) Schematically shown broadened LLs, which depict three energy scales.}
    \label{fig:qhe_gaps}
\end{figure}

By analyzing the temperature dependence of the resistance minima in the QHE regime (Fig.~\ref{fig:qhe_gaps}(a,b)), we can evaluate an energy distance $\Delta$ between the mobility edges in two subsequent LLs, according to $\sigma_{xx} = A\exp\left(-\Delta/2k_{\text{B}}T\right)$, which is plotted in Fig.~\ref{fig:qhe_gaps}(c). However, results of the \kp LLs calculations reveal that the Zeeman contribution becomes comparable with the cyclotron term at high fields \cite{Assaf2017,Krizman2018a}, and the resulting QH gaps $\Delta_{LL}$ should have almost a magnitude higher values (Fig.~\ref{fig:kp}(c)). Such discrepancy points to the large LL broadening $\Gamma$. The LL broadening can be determined through $\Gamma = \hbar/2\tau_q$ \cite{Dietl1978}, where $\tau_q$ is a quantum lifetime. The $\tau_q$ is usually determined from the Dingle analysis of the low-field SdH oscillations. In our case, we fitted low-field SdH with a sum of two cosine functions:
\begin{multline}
    \Delta R_{xx}(B) = C_{1}e^{-\pi/B\mu_{1}}\cos\left(2\pi\left(\frac{B_{F}-\Delta B_{F}}{B}+\frac12+\phi_{1}\right)\right)\\ + C_{2} e^{-\pi/B\mu_{2}} \cos\left(2\pi\left(\frac{B_{F}+\Delta B_{F}}{B}+\frac12+\phi_{2}\right)\right),
    \label{eq:LK_field}
\end{multline}
where $C_{1,2}$ are fitting constants, $\mu_{1,2}=e\tau_{q}/m^{*}$ are quantum mobilities corresponding for each frequency, $B_{F}$ is the main frequency and $\Delta B_{F}$ determines the splitting of the main frequency, finally $\phi_{1,2}$ are phase shifts for each frequency peak. After fitting the SdH oscillations at the lowest temperature, the obtained values for $B_{F}$ and $\Delta B_{F}$ were similar to those acquired from the FFT analysis. The results of the SdH fittings and the corresponding values of $\mu_{1,2}$ are shown and discussed in more detail in \cite{sup}. Quantum lifetimes have similar values for all QWs and fall to the range of 0.14--0.44~ps, corresponding to the $\Gamma\approx5$--15~meV. Thus, all three energy scales (Fig.~\ref{fig:qhe_gaps}d) are in qualitative agreement with each other.

Also, according to the \kp model (see \hyperref[app:kp]{Appendix}), for the samples with a single occupied subband, odd $\nu$ should correspond to the pure Zeeman gap which also includes exchange term ($\Delta\propto E_{z}=2g\mu_B B+E_{exch}$), while even $\nu$ -- to the difference between cyclotron and Zeeman gaps ($\Delta\propto h\omega_c - E_{z}$). While changing the tilt angle, the Zeeman contribution remains constant (increases) with $B_{tot}$ ($B_{\perp}$), while cyclotron energy reduces (remains constant) with $B_{tot}$ ($B_{\perp}$). According to such a naive picture, the energy gaps for odd $\nu$'s should decrease, and for even $\nu$'s -- increase. Nevertheless, in the $p$-type (Pb,Sn)Se QWs situation is the opposite due to the role of exchange contribution to the Zeeman energy (see \hyperref[app:kp]{Appendix}). Numerical simulations show only limited qualitative agreement between experimental and theoretical results.

\subsection{SdH fitting with \kp model}
\label{sec:kp}

\begin{figure}
    \centering
    \includegraphics[width=\columnwidth]{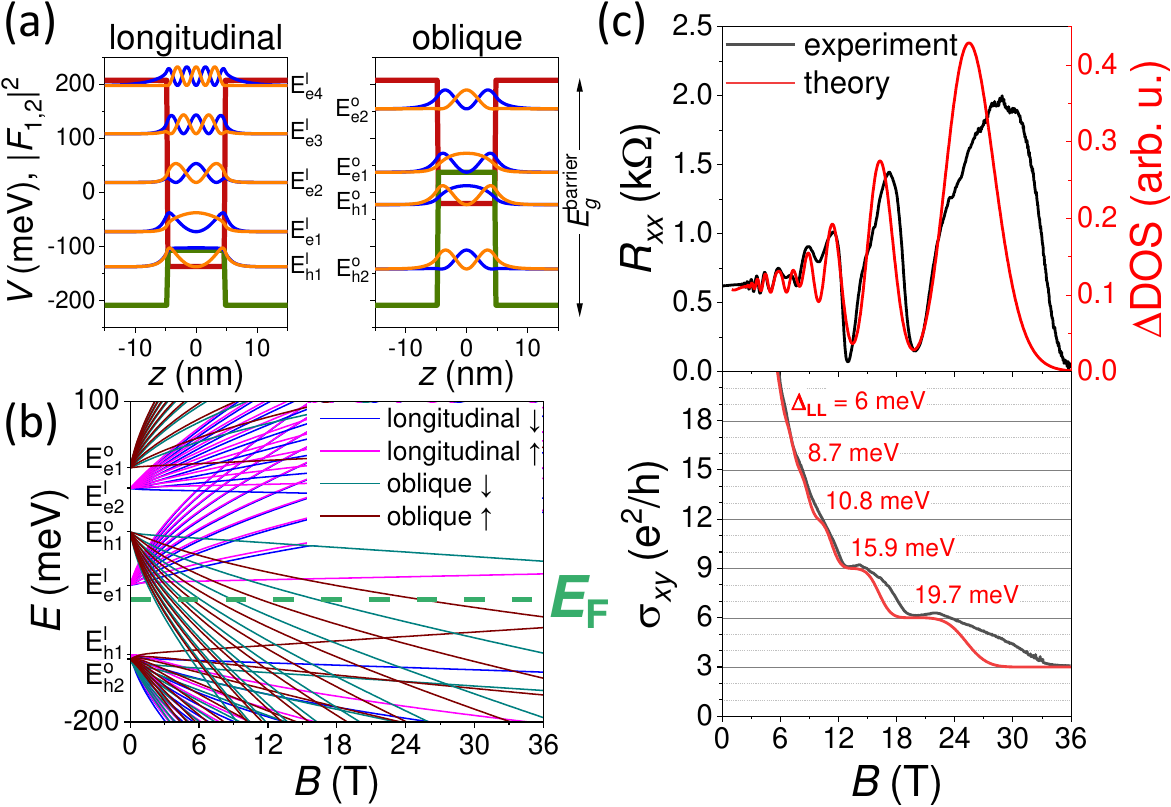}
    \caption{Results from the \kp modelling of the 10~nm QW (sample~D). (a) The potential profile of the QW is plotted independently for the longitudinal and oblique valleys. The inverted gap in the longitudinal valley is shifted toward the valence band of the barrier much stronger than the gap of the oblique valleys. This is due to the strains induced by the barriers. For clarity, the square modulus of the wavefunctions $|F_{1}|^2$ (blue) and $|F_{2}|^2$ (orange) are shifted by the values of the corresponding subband energy level. (b) First ten spin-resolved LLs as a function of the magnetic field applied perpendicular to the QW and (c) the corresponding DOS oscillations and Hall conductivity, compared to the experimental results. The theoretical curves align almost perfectly with the experimental data. The corresponding distance between nearest LLs is indicated for each filling factor, starting from $\nu = 6$.}
    \label{fig:kp}
\end{figure}

By diagonalizing the \kp Hamiltonian in zero magnetic fields, we obtained subbands' wavefunctions -- the Kramers doublet $F_{1,2}$ (\hyperref[app:kp]{Appendix}), whose squared modulus is shown in Fig.~\ref{fig:kp}(a). Here we can see that the wavefunction's tails penetrate the barriers. As already mentioned, this penetration results in strong alloy scattering reducing mobility values in thin QWs and in a non-zero $s-f$ exchange interaction between the carriers and Eu spins, similar to the Mn-doped barriers of the DMS QWs \cite{Gaj1994}. Exchange spin-splittings are crucial for the appearance of the quantum anomalous Hall effect in topological matter \cite{Chang2013,Fang2014,Chang2023}. By incorporating the exchange interactions into the \kp Hamiltonian, we are able to quantitatively describe the SdH oscillations in sample~D. The corresponding LL diagram is shown in Fig.~\ref{fig:kp}(b). The resulting LLs are non-linear, with a distance between individual LLs being of the order of 10-30~meV. At the Fermi level, $E_{F}=-85.3$~meV, we computed both density of states (DOS) oscillations and the Hall conductivity, as illustrated in Fig.~\ref{fig:kp}(c). Such $E_{F}$ lies between $E_{h1}^{l}$ and $E_{e1}^{l}$ and only $E_{h1}^{o}$ is occupied (Fig.~\ref{fig:kp_subbands}a), which is consistent with the observation of the p-type carriers and with the degeneracy of the observed QH states. The resulting DOS oscillations closely match the experimental data, highlighting the model's quantitative accuracy and its capability to provide insights into the band structure of IV-VI quantum wells under realistic experimental conditions. This means it can be used to analyse the experimental data and foresee ways to improve the heterostructure design for specific needs. However, since LL calculations require a full set of band parameters, which dependence on Sn and Eu content is not well known, this, together with other assumptions (see \hyperref[app:kp]{Appendix} for details), did not allow us to reproduce DOS oscillations in thicker QWs.

\section{Summary and Outlook}

In summary, we found several quantum features in hole transport in QWs of Sn-rich (Pb,Sn)Se with a relatively large inverted bulk band gap. The observed and quantitatively interpreted phenomena include WAL magnetoresistance, aperiodic UCF, SdH oscillations, and the quantum Hall effect. The obtained results illustrate the interplay of strain, confinement, spin-orbit coupling, exchange interaction with barrier Eu spins, and multi-valley physics in the vicinity of the topological phase transition. To interpret our findings, we numerically solved \kp Hamiltonian, which has allowed us to quantitatively reproduce the experimental data for narrow QWs. This approach has the potential to quantitatively predict and describe other magnetotransport phenomena with precision comparable to that achieved for HgTe QWs \cite{Novik2005,Bernevig2006,Beugeling2025}. Furthermore, we have found persistent photoconductivity in the barrier material at low temperatures allowing to control the carrier concentration in the (Pb,Sn)Se QWs \cite{sup}.

Our work also highlights challenges and opportunities for future progress. Though further increasing Sn content while maintaining low carrier density is needed, our results also call for improvements in the barrier material engineering. The (Pb,Eu)Se barrier material employed in most of the current studies \cite{Krizman2022,Krizman2024,Wang2023} introduces significant strain, such that the gaps of the longitudinal and oblique valleys do not coincide. While this strong partial lifting of valley degeneracy may have potential applications in valleytronics \cite{Krizman2024}, it also hinders the realization of quantized topological conduction by causing the hybridization of the helical or chiral edge channels with bulk carriers. One of the possible ways to overcome such a situation would be to engineer the barrier materials carefully. A potential candidate for the barrier material would be the Eu$_{x}$Sr$_{1-x}$S$_{y}$Se$_{1-y}$ compound \cite{Westerholt1981,Westerholt1981a}, which offers simultaneous control over lattice constant, energy gap, and magnetism in the desired range. 

In general terms, the achieved here progress in developing high quality quantum wells with a relatively large Sn and Eu content, together with the collected experimental results and theoretical modelling, constitutes a significant step for obtaining quantum structures with chemical contents, thicknesses, and strains, for which charge transport along topological edge states will dominate. In particular, according to the equilibrium phase diagram, the rock-salt structure is unstable for $x_{Sn}\geq0.4$. However, the employment of strain compensating Eu$_{x}$Sr$_{1-x}$S$_{y}$Se$_{1-y}$ barriers, as proposed above, should facilitate pseudomorphic MBE growth of cubic Pb$_{1-x}$Sn$_{x}$Se with large Sn content and, thus, with a topological gap of the order of tens meV, without precipitations of the orthorhombic phase \cite{Reddy2024}. Furthermore, those barriers offer a possibility for ferromagnetic ordering that may result in the quantum anomalous Hall effect. Considering unique properties of IV-VI compounds, such as a large magnitude of the dielectric constant, which may kill the presence of singly-occupied paramagnetic impurity centers, and a multivalley band structure that could result in large Chern numbers, this materials system offers a worthwhile prospect for quantum metrology and, after proximitizing with a superconductor, in quest for Majorana and/or non-abelian quasiparticles. Accordingly, further growth and experimental efforts are anticipated in the directions indicated here. From another viewpoint, progress in the development of IV-VI quantum structures will allow improving figures of merit of infrared lasers and detectors that do not contain critical elements, like Ga and In.

\section*{Acknowledgments}

The authors thank G\"unther Bauer and Gauthier Krizman for valuable discussions. This research was supported by the “MagTop” project (FENG.02.01-IP.05-0028/23) carried out within the “International Research Agendas” programme of the Foundation for Polish Science co-financed by the European Union under the European Funds for Smart Economy 2021-2027 (FENG). V.V.V. also acknowledges long-term program of support of the Ukrainian research teams at the Polish Academy of Sciences carried out in collaboration with the U.S. National Academy of Sciences with the financial support of external partners. Measurements at high magnetic fields were supported by LNCMI-CNRS, members of the European Magnetic Field Laboratory (EMFL) and by the Ministry of Education and Science, Poland (grant no. DIR/WK/2018/07) via its membership to the EMFL. This research was supported by funds from the state budget allocated by the Minister of Science (Polska) as part of the Polish Metrology II programme project no. PM-II/SP/0012/2024/02 amount of funding 944,900.00 PLN, total project value PLN 944,900.00.

V.V.V. grew the heterostructures with the assistance of G.S. and carried out structural characterization; A.K. carried out microstructure processing and all magnetotransport measurements with the assistance of Z.A.; C.-W.C. and B.A.P. assisted in high-field magnetotransport measurements. T.~Wojciechowski performed an EDX analysis and assisted in microstructure processing. A.K. and T.D. performed \kp calculations. T.~Wojtowicz and T.D. were responsible for funding acquisition and general management. The manuscript was written by A.K., V.V.V., B.A.P., and T.D. All authors discussed the results and commented on the manuscript.

\section*{Data availability}

The data that support the findings of this article are openly available \cite{Kazakov_2025}.

{
\appendix*

\begin{widetext}
\section{\kp model}
\label{app:kp}

\subsection{Low magnetic fields}
We start with the four band \kp Hamiltonian $\hat{H}$ \cite{Mitchell1966,Nimtz1983,Bauer1992}:
\begin{equation}
    \begin{pmatrix}
        \begin{matrix}V_v(z)-\frac{\hbar^2}{2m^v_t}(k_x^2+k_y^2)\\-\frac{\hbar^2}{2m_l^v}k_z^2-\frac12g_l^v\mu_BB_z\end{matrix} & -\frac12g_t^v\mu_B(B_x-iB_y) & \hbar v_ck_z & \hbar v_c(k_x-ik_y) \\
        -\frac12g_t^v\mu_B(B_x+iB_y) & \begin{matrix}V_v(z)-\frac{\hbar^2}{2m^v_t}(k_x^2+k_y^2)\\-\frac{\hbar^2}{2m_l^v}k_z^2+\frac12g_l^v\mu_BB_z\end{matrix} & \hbar v_c(k_x+ik_y) & -\hbar v_ck_z \\
        \hbar v_ck_z & \hbar v_c(k_x-ik_y) & \begin{matrix}V_c(z)+\frac{\hbar^2}{2m^c_t}(k_x^2+k_y^2)\\+\frac{\hbar^2}{2m_l^c}k_z^2+\frac12g_l^c\mu_BB_z\end{matrix} & \frac12g_t^v\mu_B(B_x-iB_y) \\
        \hbar v_c(k_x+ik_y) & -\hbar v_c k_z & \frac12g_t^v\mu_B(B_x+iB_y) & \begin{matrix}V_c(z)+\frac{\hbar^2}{2m^c_t}(k_x^2+k_y^2)\\+\frac{\hbar^2}{2m_l^c}k_z^2-\frac12g_l^c\mu_BB_z\end{matrix} \\
    \end{pmatrix}
\end{equation}
Here, $k_z=-i\hbar\partial/\partial z$ is aligned along the growth direction, i.e., [111]; $B_x$, $B_y$, $B_z$ give Zeeman splitting with respect to the tilted magnetic field. Other band parameters are listed in Table~\ref{tab:bandparam} and taken as constants throughout the heterostructure. We assume that changes in the band parameters over the interface are negligible because of (i) a small level of Eu- and Sn- alloying relative to the parent PbSe compound and (ii) the low penetration amplitude of the resulting wavefunctions into the barrier. However, we used a slightly different set of band parameters for the longitudinal and oblique valleys.

Additionally, we also introduce the $sp$-$f$ exchange interaction between band carriers and Eu ions residing in the barriers. We consider only the exchange term for holes, as exchange coupling is small for electrons \cite{Bauer1992,Dietl1994}. We added the following term to the initial Hamiltonian:
\begin{equation}
    \begin{pmatrix}
        \Delta_{sf}*w_{z}/2 & \Delta_{sf}*\left(w_{x}-iw_{y}\right)/2 & 0 & 0 \\
        \Delta_{sf}*\left( w_{x}+iw_{y}\right)/2 & -\Delta_{sf}*w_{z}/2 & 0 & 0 \\
        0 & 0 & 0 & 0 \\
        0 & 0 & 0 & 0 \\
    \end{pmatrix}
\end{equation}
Here, $w_i$ are versors describing the direction of the total magnetic field $B$; $\Delta_{sf}=y_{\text{Eu}}AN_{0}S\mathcal{B}_{s}(x_{B})$, where $y_{\text{Eu}}$ is the Eu content in the barriers, $AN_{0}=80$~meV is the exchange energy, and $S=7/2$ is Eu spin. $\mathcal{B}_{s}(x_{B})$ is the Brillouin function with an argument $x_{B}=Sg_{\text{Eu}}\mu_{B}B/k_{B}T$, where $g_{\text{Eu}}=2$ is the Eu spin $g$-factor.

We assume band offsets $V_{c,v}(z)$ to change across the (Pb,Sn)Se/(Pb,Eu)Se interface according to,
\begin{equation}
    V_{c,v}(z)=
    \begin{cases}
        \pm E_{\text{gap}}^{\text{barrier}}, \text{ if } |z|>d_{\text{QW}}\\
        \pm E_{\text{gap}}^{\text{QW}}+E_{\text{strain}}, \text{ if } |z|\leq d_{\text{QW}}
    \end{cases},
\end{equation}
where '$+$' sign is chosen for conduction ($V_c$) bands and '$-$' -- for the valence ($V_v$) bands. Here, $E_{\text{gap}}^{\text{QW}}$ is the energy gap of the QW material, and $E_{\text{gap}}^{\text{barrier}}$ of the barrier material. QW energy gap is determined from \cite{Nimtz1983}:
\begin{equation}
E_{\text{gap}}^{\text{QW}}(x_{\text{Sn}}, T)[\text{meV}] = 125 - 1021x_{\text{Sn}} + \sqrt{400+0.256*T[\text{K}]^2},
\end{equation}
and the barrier energy gap is determined from \cite{Krizman2018}:
\begin{equation}
E_{\text{gap}}^{\text{barrier}}(x_{\text{Eu}}, T)[\text{meV}] = (146 + \frac{1 - 3 x_{\text{Eu}}}{T[\text{K}] + 40.7} 0.475*T[\text{K}]^2 + 3000*x_{\text{Eu}}).
\end{equation}

Here we also include the effect of strain on the band structure \cite{Zasavitskii2004}, by considering the $E_{\text{strain}}$ term. We assume that the barriers are completely relaxed and the strain in the QW is determined by the ratio between the corresponding lattice constants \cite{Zasavitskii2004,Simma2014}:
\begin{equation}
    \varepsilon_{\|}= 1 - a^{\text{QW}} / a^{\text{barrier}},
\end{equation}
where $a^{\text{QW}}$ is lattice constant of unstrained QW material \cite{Krizman2018a}:
\begin{equation}
    a^{\text{QW}}(x_{\text{Sn}}) = 6.124 - 0.1246*x_{\text{Sn}},
\end{equation}
and $a^{\text{barrier}}$ is the lattice constant of the barrier material \cite{Krizman2024}:
\begin{equation}
    a^{\text{barrier}}(y_{\text{Eu}}) = 6.124 + 0.025*y_{\text{Eu}}
\end{equation}
Since here comes the ratio between lattice constants, we do not consider the effect of the thermal expansion following the reasoning provided in \cite{Simma2014}. Then, the out-of-plane strain is \cite{Zasavitskii2004}:
\begin{equation}
    \varepsilon_{\perp}=-2\frac{C_{11}+2C_{12}-2C_{44}}{C_{11}+2C_{12}+4C_{44}}\varepsilon_{\|},
\end{equation}
where elastic constants $C_{11}$, $C_{12}$, and $C_{44}$ are 14.18, 1.94, and 1.749, respectively \cite{Zasavitskii2004}. The strain shifts the band edges of the material differently for longitudinal and oblique valleys \cite{Simma2014}:
\begin{gather}
    \delta E^{c,v}_{l}=D^{c,v}_d\left( 2\varepsilon_{\|}+\varepsilon_{\perp}\right) + D^{c,v}_u\varepsilon_{\perp}\\
    \delta E^{c,v}_{o}=D^{c,v}_d\left( 2\varepsilon_{\|}+\varepsilon_{\perp}\right) + D^{c,v}_u\left( 8\varepsilon_{\|}+\varepsilon_{\perp}\right)/9
\end{gather}

Values for the deformation potentials $D^{c,v}_{d,u}$ were taken from \cite{Zasavitskii2004} and slightly adjusted to better match the published experimental data on the dispersion and oblique/longitudinal valley offsets \cite{Rechcinski2021,Krizman2024}, Table \ref{tab:deform_pot}. Thus, to account for the strain effect, $\delta E^{c,v}_{l,o}$ substitutes $E_{\text{strain}}$ and the corresponding Schr\"odinger problem:
\begin{equation}
    \hat{H}\vec{f_n}=E_n\vec{f_n}
\end{equation}
is solved separately for the longitudinal and oblique valleys. The eigenvector $\vec{f_n}$ is expanded as:
\begin{equation}
    \vec{f_n}=\sum_{j=-m}^{m}c^{j}_n \exp{ \left( i\frac{2\pi nz}{L_z} \right) }/\sqrt{L_z},
\end{equation}
where $m$ is between 50 and 100 to ensure an appropriate numerical convergence, $L_z=d_{\text{QW}}+\mathrm{100~nm}$ is the total thickness of the considered structure.

At zero magnetic field ($B=0$) and zero momentum ($k_{x}=k_{y}=0$), $\hat{H}$ can be reduced to two $2\times2$ eigenvalue problems:

\begin{equation}
\begin{aligned}
        \begin{pmatrix}
        V_v(z)-\frac{\hbar^2}{2m_l^v}k_z^2 & \hbar v_ck_z \\
        \hbar v_ck_z & V_c(z)+\frac{\hbar^2}{2m_l^c}k_z^2 \\
    \end{pmatrix}
    \begin{pmatrix} F_{1}^{i}(z) \\ F_{2}^{i}(z) \\ \end{pmatrix}
    & = E_{i} \begin{pmatrix} F_{1}^{i}(z) \\ F_{2}^{i}(z) \\ \end{pmatrix} \\
    \begin{pmatrix}
        V_v(z)-\frac{\hbar^2}{2m_l^v}k_z^2 & -\hbar v_ck_z \\
        -\hbar v_c k_z & V_c(z)+\frac{\hbar^2}{2m_l^c}k_z^2 \\
    \end{pmatrix}
    \begin{pmatrix} F_{1}^{i}(z) \\ -F_{2}^{i}(z) \\ \end{pmatrix}
    & = E_{i} \begin{pmatrix} F_{1}^{i}(z) \\ -F_{2}^{i}(z) \\ \end{pmatrix}
    \label{eq:subbands}
\end{aligned}
\end{equation}

Here, $F_{1}^{i}(z)$ and $F_{2}^{i}(z)$ are valence $L_6^v$ and conduction $L_6^c$ components of the wavefunction for the $i$'th subband. Solutions of eq.~\ref{eq:subbands} were used to calculate subbands energy levels (Fig.~\ref{fig:kp_subbands}) and model the phase diagram (Fig.~\ref{fig:kp_th}).

\begin{table}
    \centering
    \caption{Parameters of longitudinal (l) and oblique (o) valleys used in the calculations. Note that compared to the notations used in \cite{Mitchell1966,Nimtz1983}, $P$ is replaced with the Fermi velocity $v_f$ \cite{Krizman2018}: $v_f=P/m_0$, with $m_0$ being the electron rest mass. The masses are given in electron mass units.}
    \begin{tabular}{|c|c c c c c c c c c|}
        \hline
        valley & $m^v_l$ & $m^v_t$ & $m^c_l$ & $m^c_t$ & $g^v_l$ & $g^v_t$ & $g^c_l$ & $g^c_t$ & $v_f$ [m/s]\\
        \hline
        l & 0.37 & 0.20 & 0.37 & 0.20 & -3.3 & -0.8 & -5.1 & -3.5 & $4.5\times10^5$  \\
        \hline
        o & 0.08 & 0.09 & 0.08 & 0.09 & -3.89 & -0.94 & -6.01 & -4.13 & $4.87\times10^5$  \\
        \hline
    \end{tabular}
    \label{tab:bandparam}
\end{table}

\begin{table}
    \centering
    \caption{Values of the deformation potentials, used in the calculations.}
    \begin{tabular}{c c c c }
        \hline
        $D_d^c$ & $D_u^c$ & $D_d^v$ & $D_u^v$ \\
        \hline
        -4.36 & 8.29 & -8.93 & 10.46 \\
        \hline
    \end{tabular}
    \label{tab:deform_pot}
\end{table}

\subsection{High magnetic fields}

To account for the quantizing magnetic fields, we substitute $k_x$ and $k_y$:
\begin{gather}
    k_x - ik_y = \sqrt{\frac{2eB}{\hbar}}a\\
    k_x + ik_y = \sqrt{\frac{2eB}{\hbar}}a^\dagger,
\end{gather}
where $a$ and $a^\dagger$ are annihilation and creation operators for the
harmonic oscillator $\phi_N$, thus $a^\dagger a\phi_N=N\phi_N$. This leads to the following Hamiltonian $\hat{H}$:

\begin{equation}
    \begin{pmatrix}
        \begin{matrix}V_v(z)-\left( N-\frac12\right)\hbar\tilde{\omega}^c\\-\frac{\hbar^2}{2m_l^v}k_z^2-\frac12g_l^v\mu_BB_z\end{matrix} & -\frac12g_t^v\mu_B(B_x-iB_y) & \hbar v_ck_z & \sqrt{2e\hbar v_c^2BN} \\

        -\frac12g_t^v\mu_B(B_x+iB_y) & \begin{matrix}V_v(z)-\left( N+\frac12\right)\hbar\tilde{\omega}^c\\-\frac{\hbar^2}{2m_l^v}k_z^2+\frac12g_l^v\mu_BB_z\end{matrix} & \sqrt{2e\hbar v_c^2BN} & -\hbar v_ck_z \\

        \hbar v_ck_z & \sqrt{2e\hbar v_c^2BN} & \begin{matrix}V_c(z)+\left( N-\frac12\right)\hbar\tilde{\omega}^v\\+\frac{\hbar^2}{2m_l^c}k_z^2+\frac12g_l^c\mu_BB_z\end{matrix} & \frac12g_t^v\mu_B(B_x-iB_y) \\

        \sqrt{2e\hbar v_c^2BN} & -\hbar v_c k_z & \frac12g_t^v\mu_B(B_x+iB_y) & \begin{matrix}V_c(z)+\left( N+\frac12\right)\hbar\tilde{\omega}^v\\+\frac{\hbar^2}{2m_l^c}k_z^2-\frac12g_l^c\mu_BB_z\end{matrix} \\
    \end{pmatrix},
\end{equation}
where $\tilde{\omega}^{c,v}=eB/\tilde{m}^{c,v}_t$ is the cyclotron frequency. We have not considered here the effect of the in-plane magnetic field mixing spatial and momentum variables \cite{Winkler2003}. Then, the Landau levels are calculated up to $N$ = 30 using the same approach as described above. Note that the $N=0$ LL is spin polarized in the case of an inverted gap since the topological surface states form it \cite{Assaf2017}.

The oscillations of the density of states (DOS) at the Fermi level $E_F$ are estimated through:
\begin{equation}
    \mathrm{DOS}\propto \sum_n\exp\left( -\frac{(E_F-E_n^{\text{LL}})}{\Gamma}\right),
\end{equation}
where $\Gamma$ describes Landau-level broadening.

Such a simplified picture nevertheless captures the general trend of the magnetic field tilt angle dependence of the SdH oscillations. In Fig.~\ref{fig:app_kp_tilt} we compared experimental results (upper panel) with the outcome of numerical modelling (lower panel). Indeed, according to the naive picture, the energy gaps for the even $\nu$'s should increase with a tilt angle as a function of $B_{\perp}$, since they are purely Zeeman gaps. This is not the case either for experimentally measured curves or for the results of the \kp calculations. Further investigation reveals that the faster closing of the Zeeman gap with increasing tilt angle is primarily driven by exchange interactions. Nonetheless, a more detailed and comprehensive model is required to fully explain this behavior.
\end{widetext}

\begin{figure}
    \centering
    \includegraphics[width=0.80\columnwidth]{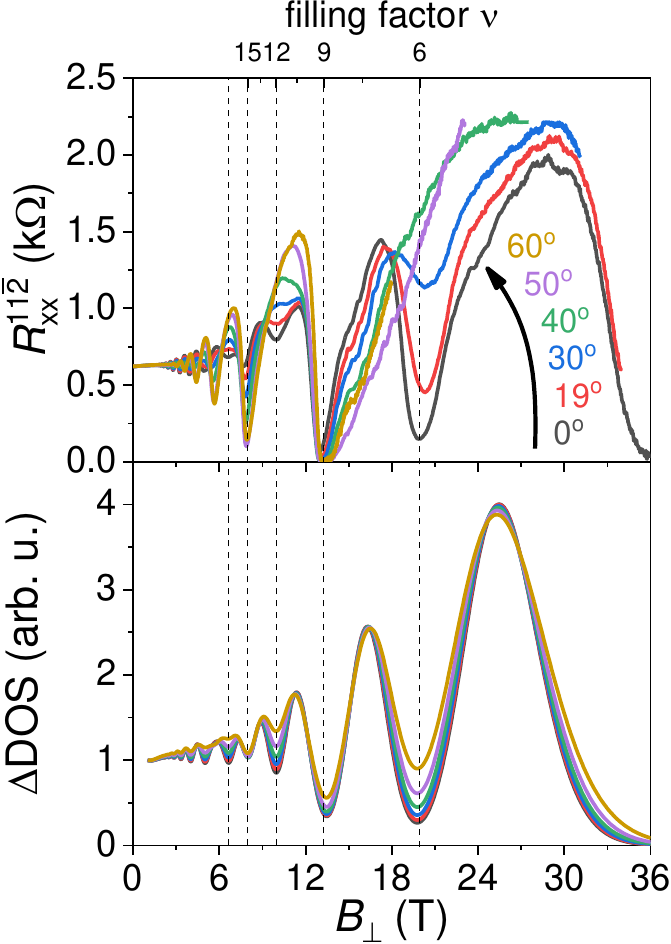}
    \caption{Comparison of the evolution of the SdH and QHE with a tilt angle for sample~D obtained from the experiment (upper panel) and \kp calculations (lower panel).}
    \label{fig:app_kp_tilt}
\end{figure}
}

\end{document}


\author{Alexander~Kazakov\orcidA}
\email{kazakov@MagTop.ifpan.edu.pl}
\affiliation{International Research Centre MagTop, Institute of Physics, Polish Academy of Sciences, Aleja Lotnikow 32/46, PL-02668 Warsaw, Poland}
\author{Valentine~V.~Volobuev\orcidH}
\email{volobuiev@MagTop.ifpan.edu.pl}
\affiliation{International Research Centre MagTop, Institute of Physics, Polish Academy of Sciences, Aleja Lotnikow 32/46, PL-02668 Warsaw, Poland}
\affiliation{National Technical University "KhPI", Kyrpychova Str. 2, 61002 Kharkiv, Ukraine}
\author{Chang-Woo~Cho\orcidC}
\affiliation{Laboratoire National des Champs Magn{\'e}tiques Intenses, CNRS, LNCMI, Universit{\'e} Grenoble Alpes, Universit{\'e} Toulouse 3, INSA Toulouse, EMFL, F-38042 Grenoble, France}
\affiliation{Department of Physics, Chungnam National University, Daejeon, 34134, Republic of Korea}
\author{Benjamin~A.~Piot\orcidD}
\affiliation{Laboratoire National des Champs Magn{\'e}tiques Intenses, CNRS, LNCMI, Universit{\'e} Grenoble Alpes, Universit{\'e} Toulouse 3, INSA Toulouse, EMFL, F-38042 Grenoble, France}
\author{Zbigniew~Adamus\orcidB}
\affiliation{Institute of Physics, Polish Academy of Sciences, Aleja Lotnikow 32/46, PL-02668 Warsaw, Poland}
\author{Tomasz~Wojciechowski\orcidE}
\affiliation{International Research Centre MagTop, Institute of Physics, Polish Academy of Sciences, Aleja Lotnikow 32/46, PL-02668 Warsaw, Poland}
\author{Tomasz~Wojtowicz\orcidG}
\affiliation{International Research Centre MagTop, Institute of Physics, Polish Academy of Sciences, Aleja Lotnikow 32/46, PL-02668 Warsaw, Poland}
\author{Gunther~Springholz\orcidF}
\affiliation{Institut f{\"u}r Halbleiter- und Festk{\"o}rperphysik, Johannes Kepler University, Altenbergerstrasse 69, A-4040 Linz, Austria}
\author{Tomasz~Dietl\orcidI}
\email{dietl@MagTop.ifpan.edu.pl}
\affiliation{International Research Centre MagTop, Institute of Physics, Polish Academy of Sciences, Aleja Lotnikow 32/46, PL-02668 Warsaw, Poland}

\onecolumngrid
\renewcommand{\thefigure}{S\arabic{figure}}
\renewcommand{\thetable}{S\arabic{table}}
\renewcommand{\theequation}{S\arabic{equation}}
\renewcommand{\thepage}{S\arabic{page}}
\renewcommand{\thesection}{S\arabic{section}}

\setcounter{page}{1}
\setcounter{section}{0}
\setcounter{equation}{0}
\setcounter{figure}{0}
\setcounter{table}{0}

\begin{center}
\textbf{\Large Supplemental Material} \\
\vspace{0.2in} \textmd{\Large \thetitle}\\
\end{center}

\section{Characterization of unprocessed quantum wells}

\begin{figure}[hb]
    \centering
    \includegraphics[width=0.70\columnwidth]{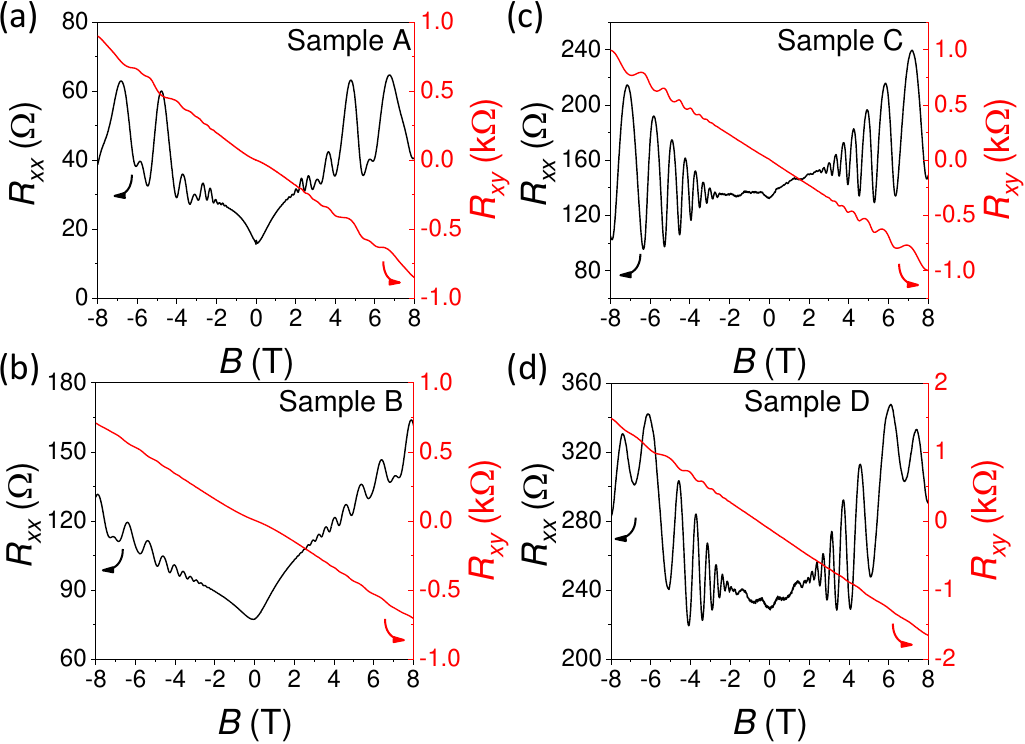}
    \caption{Magnetoresistance (black) and Hall resistance (red) for unprocessed QW samples~A-D at 1.5~K.}
    \label{fig:sup_asgrown}
\end{figure}

The results of the initial characterization of the studied QWs are presented in Fig.~\ref{fig:sup_asgrown}. Previously, it was shown that lithographical processing deteriorates transport properties of the topological materials \cite{Bendias2018,Andersen2023a}. In the current samples, we did not observe that it is a substantial effect.

\begin{figure}[hb]
    \centering
    \includegraphics[width=0.70\columnwidth]{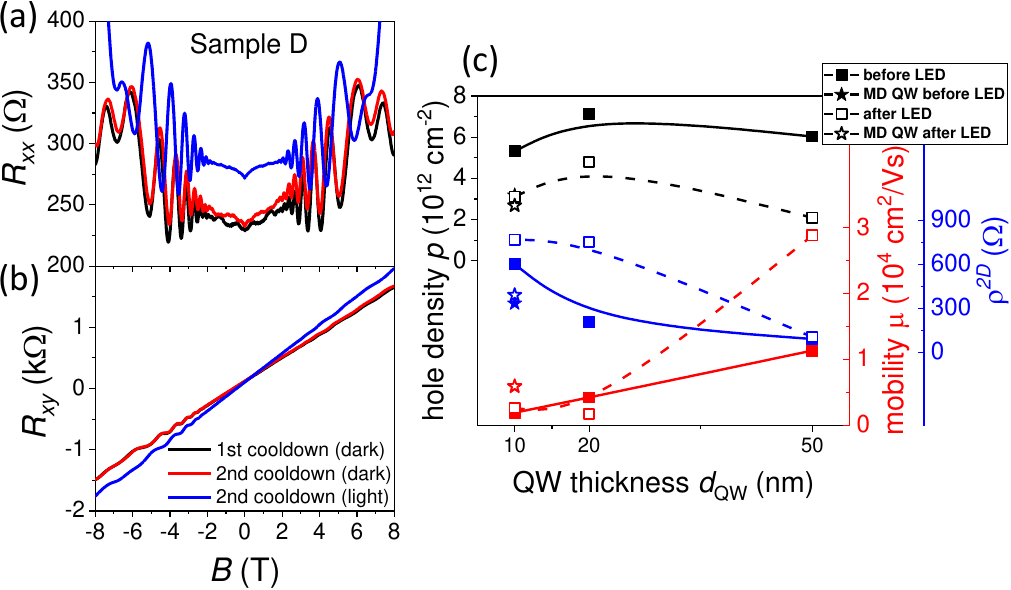}
    \caption{(a, b) Effects of illumination and thermal cycling to room temperature on longitudinal and Hall magnetoresistances at 1.6~K in sample~D. (c) Change of magnetotransport parameters in the studied samples after illumination with a green LED.}
    \label{fig:sup_thermo_light}
\end{figure}

Previously, it was found that repeated exposure to temperature variations causes structural flaws to develop in IV-VI compound layers due to differences in their expansion properties compared to the substrate. This leads to an increase in carrier density and degradation in the material's electrical transport characteristics \cite{Olver1994}. To assess the impact of thermal cycling on transport properties, we compared two successive cooldowns (Fig.~\ref{fig:sup_thermo_light}(a,b), black and red curves). We found that thermal cycling has a negligible effect on carrier density and slightly deteriorates mobility. The curves in Fig.~\ref{fig:sup_thermo_light}(a,b) were measured on unprocessed samples, however, the same behavior was observed in the Hall-bar samples, discussed in the main text.

Another common phenomenon in 2D semiconductor heterostructures is the persistent photoconductivity (PPC) effect. Indeed, such an effect was observed in the IV-VI heterostructures \cite{Pena2017,Bolanos2022}, where significant changes in the carrier density were reported. In our study of (Pb,Sn)Se/(Pb,Eu)Se QWs, we also observe the PPC effect; see red and blue lines on Fig.~\ref{fig:sup_thermo_light}(a,b). The measurement protocol has been as follows: starting from 1.6~K, the sample was warmed up to $\sim$20~K and a green LED was turned on with a 5~$\mu$A current for 1 hour until $R_{xx}$ stabilized. After that, the LED was turned off, and the sample remained for $\sim$1~hour in the dark. The sample was then cooled back down to 1.6~K, and magnetotransport measurements were taken. The hole density decreased from 3.28 to 2.67~10$^{12}$~cm$^{-2}$, while the mobility increased slightly by $\sim$1.5~\%. According to previous studies \cite{Pena2017}, the PPC effect in PbTe/(Pb,Eu)Te QWs originates from the excitation of the charges from the trap levels associated with Eu atoms. However, the role of interfacial defects cannot be excluded. It is natural to expect that the origin of the PPC effect in the selenide QWs is the same as in the telluride QWs. Thus, we also reveal that carrier density can be persistently changed by illumination at low temperatures, an important tool -- in addition to electrostatic gating -- for probing 2D or 1D edge states in QWs. Detailed study of PPC effect in PbSnSe QWs will be published elsewhere.

\clearpage
\newpage
\section{Additional data for the SdH oscillations}

\begin{figure}[hb]
    \centering
    \includegraphics[width=0.66\columnwidth]{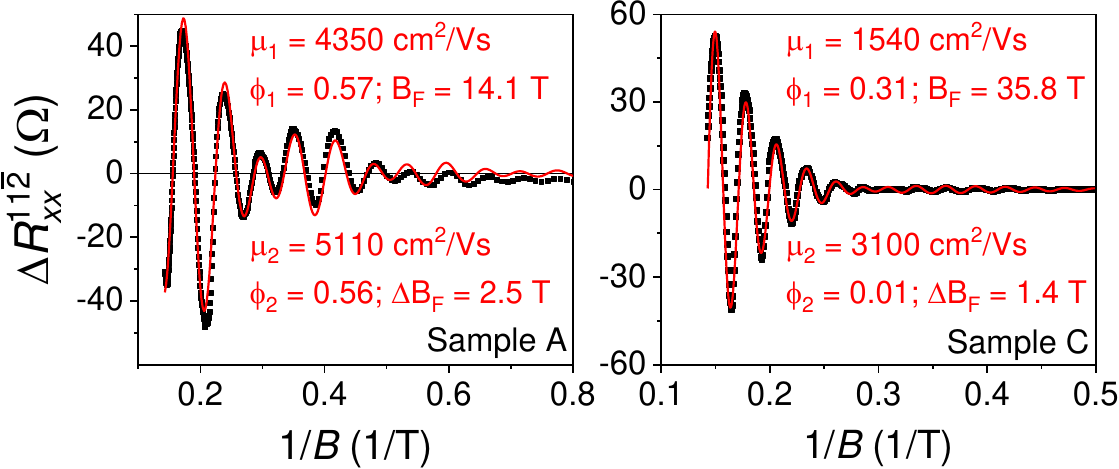}
    \caption{Results of the 2-sine function (eq.~\ref{eq:LK_field}) (main text) fitting of the SdH oscillations, measured along $[11\overline{2}]$ direction, in samples~A and C.}
    \label{fig:sup_phase}
\end{figure}

\begin{figure}[hb]
    \centering
    \includegraphics[width=\columnwidth]{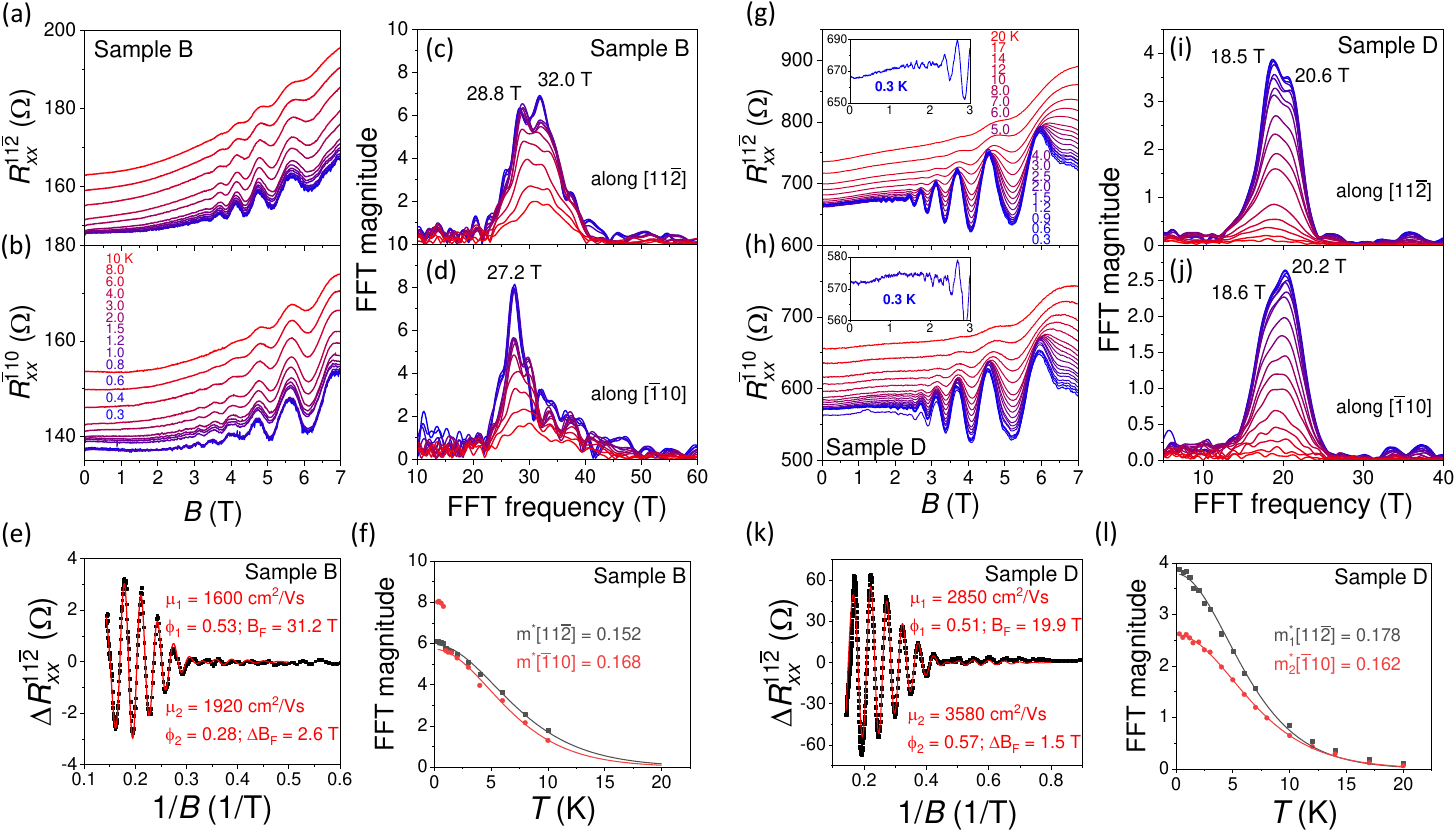}
    \caption{Additional data of the SdH oscillations in samples B (a-f) and D (g-l).}
    \label{fig:sup_sdh}
\end{figure}

Results of the two-cosine fitting of the SdH oscillations in samples A and C (presented in the main text) are shown in Fig.~
\ref{fig:sup_phase}. Additional data on the low-field ($<7$~T) SdH oscillations for samples~B (Fig.~\ref{fig:sup_sdh}(a-f)) and D (Fig.~\ref{fig:sup_sdh}(g-l)). These include raw $R_{xx}^{11\overline{2}}$ and $R_{xx}^{\overline{1}10}$ measurements, as well as FFT analysis, cyclotron $m^*$ fitting and two-cosine fitting of the $\Delta R_{xx}^{11\overline{2}}$.

\clearpage
\newpage
\section{Tilt angle dependence in moderate fields}

\begin{figure}[hb]
    \centering
    \includegraphics[width=\columnwidth]{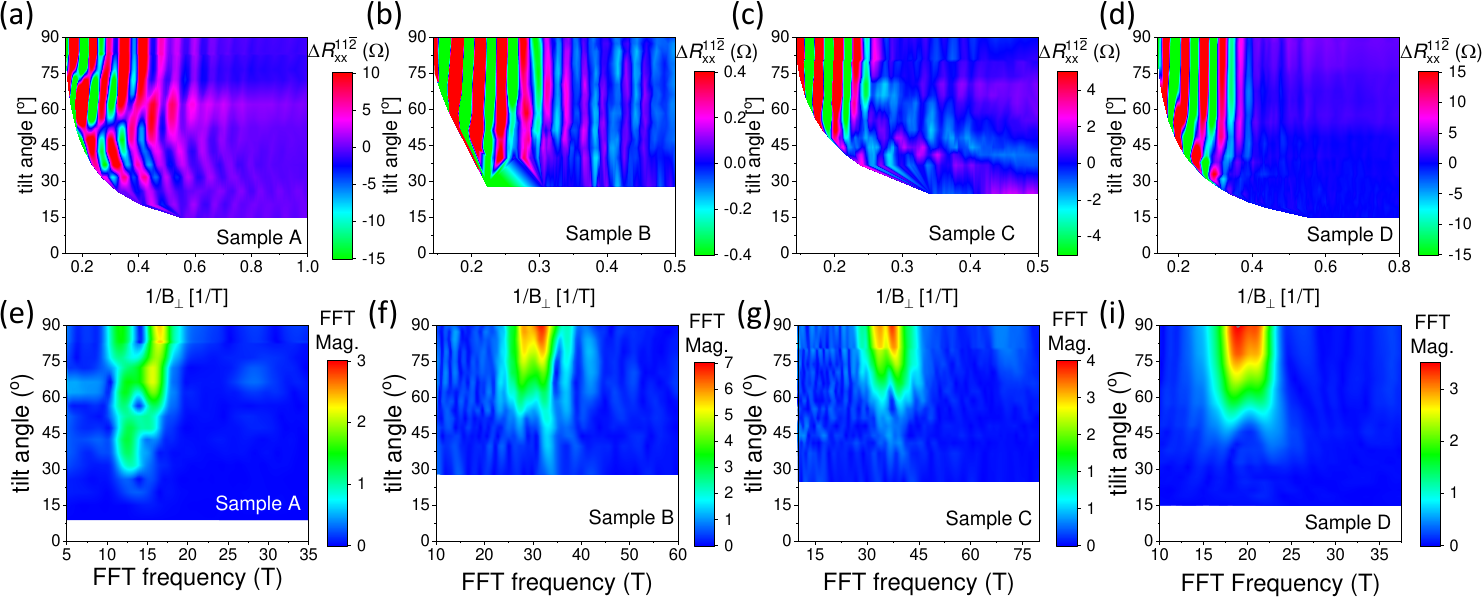}
    \caption{(a-d) Evolution of the SdH pattern, measured along $[11\overline{2}]$, with the tilt angle in the 50 and 10 nm QWs. SdH pattern is plotted vs. B$_\perp$, thus it doesn't change in a 10~nm QW sample. Note that the SdH pattern changes with the tilt angle in a 50~nm QW. (e-i) Corresponding FFT spectra performed for the $\Delta R_{xx}^{112}$(B$_\perp$), which show that the fundamental frequency $B_{F}$ is practically constant in the full range of tilt angles.}
    \label{fig:sup_sdh_tilt}
\end{figure}

In 2D (3D) systems, the frequency of SdH oscillations remains unchanged as a function of $B_{\perp}$ ($B_{\text{tot}}$). However, variations in the SdH pattern can occur due to changes in the ratio between cyclotron and Zeeman energies in 2D systems or due to band anisotropy in 3D systems. We have performed tilted field experiments for all the studied QWs (Fig.~\ref{fig:sup_sdh_tilt}). We find that the main frequency $B_{F}$ remained constant as a function of a tilt angle, which is consistent with a 2D nature of the magnetotransport of the QWs. Meanwhile, we have observed that the field tilting changes the $\Delta B_{F}$ value in some samples.

\clearpage
\newpage
\section{Additional data for the QHE}

\begin{figure}[hb]
    \centering
    \includegraphics[width=0.45\columnwidth]{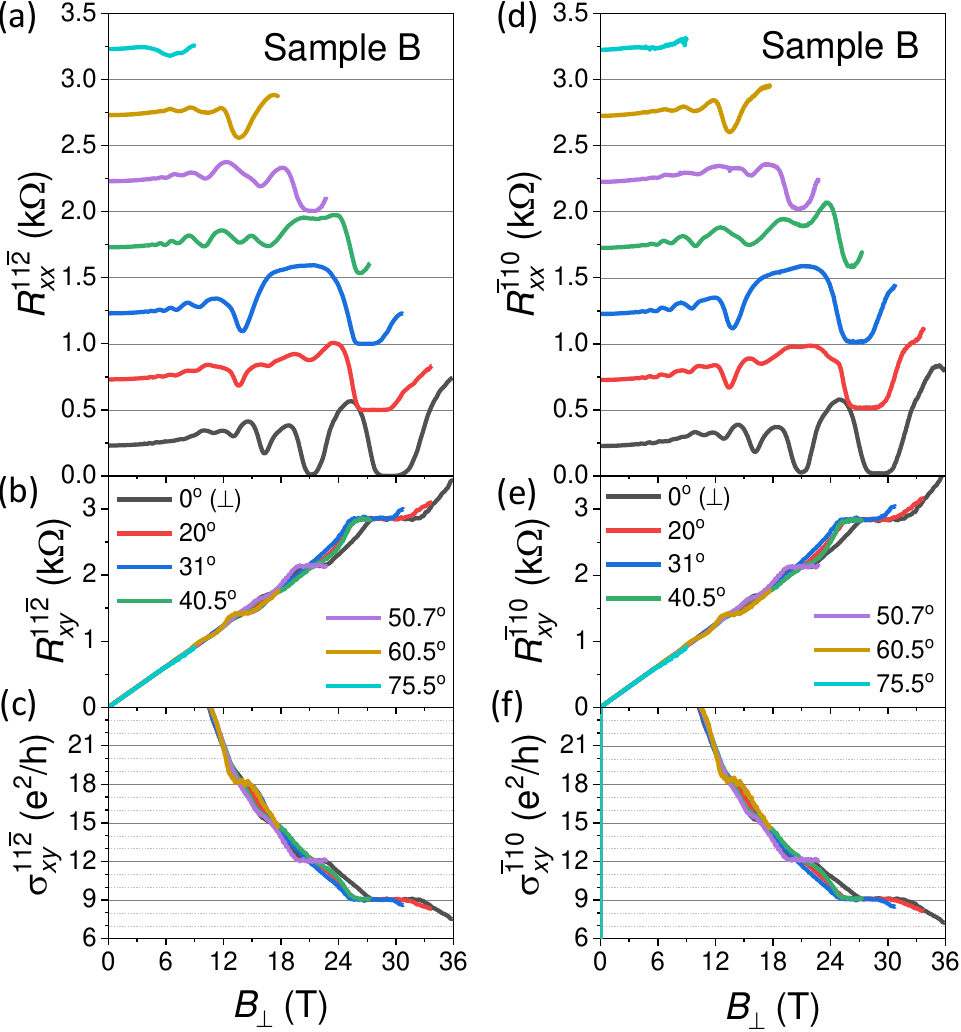}
    \caption{Additional data of the QHE in sample~B.}
    \label{fig:sup_qhe20}
\end{figure}

\begin{figure}[hb]
    \centering
    \includegraphics[width=0.90\columnwidth]{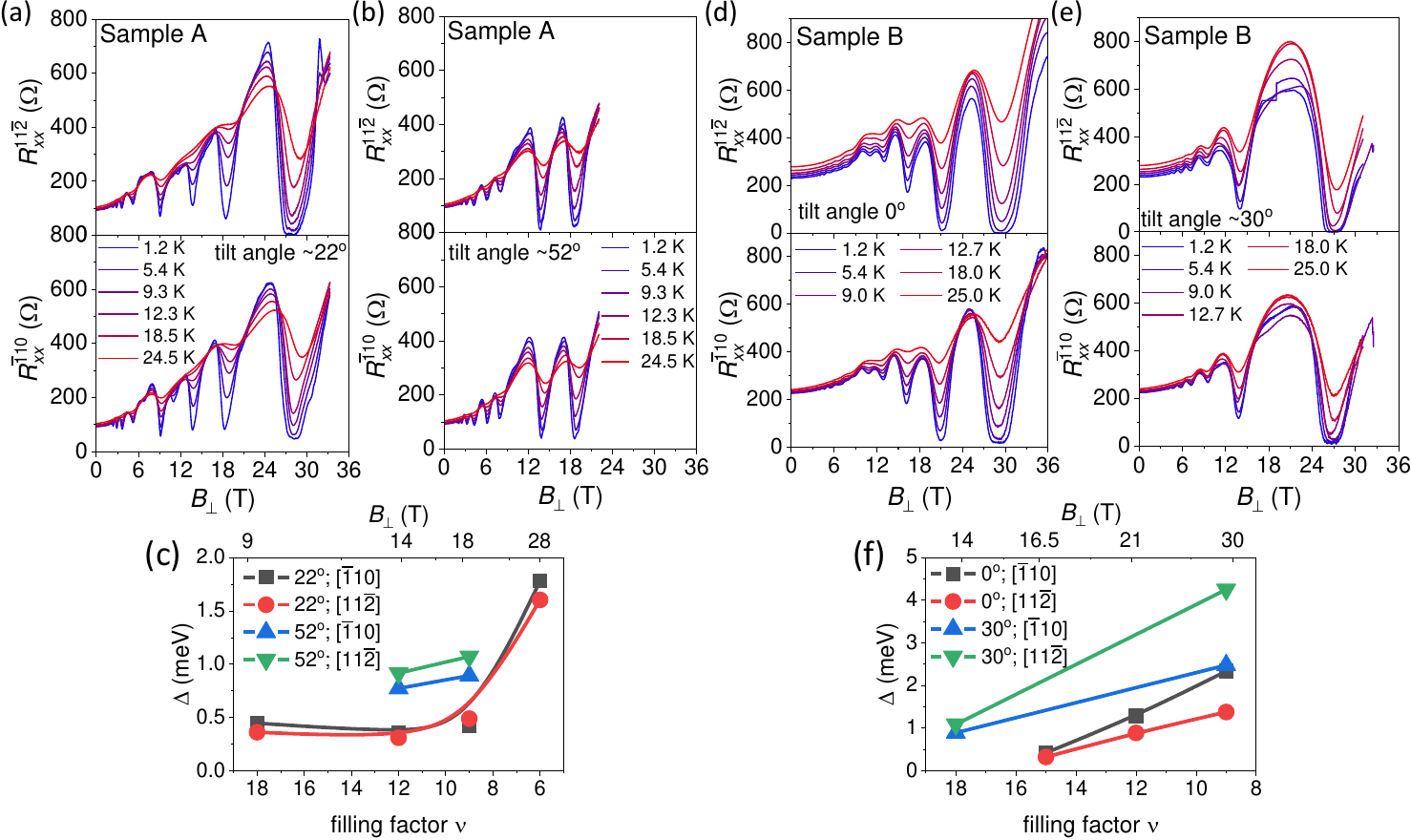}
    \caption{Additional data of the QHE energy gaps in the samples A (a-c) and B (d-e).}
    \label{fig:sup_qhe_gaps}
\end{figure}

The section contains data on the high-field QHE for samples~A (Fig.~\ref{fig:sup_qhe20} and \ref{fig:sup_qhe_gaps}) and B (Fig.~\ref{fig:sup_qhe_gaps}). These include raw $R_{xx}^{11\overline{2}}$ and $R_{xx}^{\overline{1}10}$ measurements, as well as temperature evolution of the QHE states at two tilt angles and results of their temperature damping analysis that gives an estimation of the thermal activation gap $\Delta$.

\clearpage
\newpage
\section{QW's in a parallel magnetic field}

\begin{figure}[hb]
    \centering
    \includegraphics[width=0.4\columnwidth]{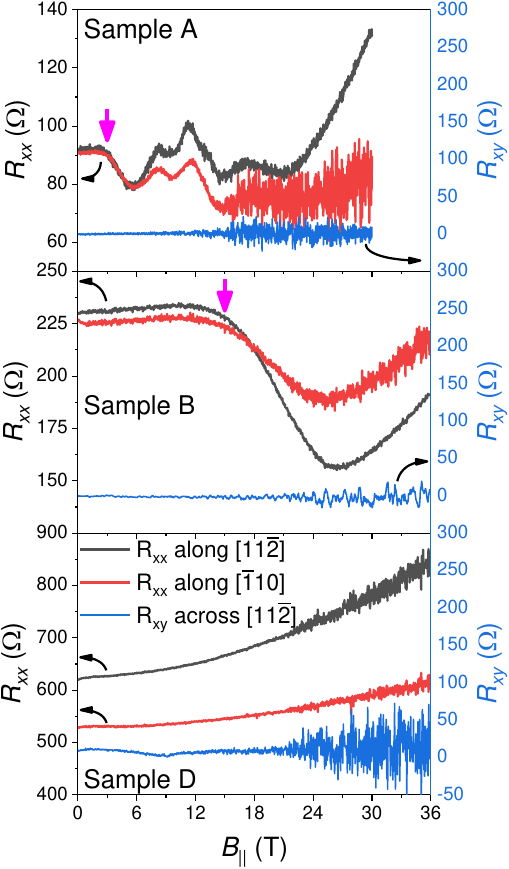}
    \caption{Magnetoresistance of three QWs as a function of the in-plane magnetic field. We also plot Hall resistance (blue), which shows that within the noise level, the magnetic field is indeed parallel to the sample plane. The magenta arrow indicates a magnetic field, which is compared with the QW width.}
    \label{fig:sup_Bpar}
\end{figure}

Being a 2D system, the quantum well should not exhibit oscillations in the parallel field. As for low magnetic fields, we have shown that this is the case. However, for high magnetic fields, when the magnetic length $l_{m}=\sqrt{\hbar/eB}$ becomes comparable or lower than the QW width, the 2D gas may exhibit quantum oscillations, due to the transition from a quantized QW subbands to LL quantization \cite{Yoshino1984}. Indeed, if we consider QW behavior in 36~T in-plane magnetic fields, we see resistance oscillations for samples~A and B (Fig.\ref{fig:sup_Bpar}). In low magnetic fields of several Teslas, we see small positive magnetoresistance, which is interrupted with abrupt turndown. We can compare the onset of this turndown in magnetic length units with the QW's thickness. This onset is $\approx$~3~T and $\approx$~15~T for sample~A and B, respectively. These magnetic fields correspond to $\approx$~14.8~nm and $\approx$~6.6~nm, about 1/3 of the QW width for both thicknesses. For sample~D, this ratio of 1/3 gives an estimation of $\approx$60~T for the onset of the oscillations, which lies beyond the range of fields employed within our study.

